\documentclass[twocolumn]{autart}

\usepackage{graphicx}     
\usepackage{url} 
\usepackage{algorithm}
\usepackage{algorithmic}
\usepackage{amsmath,amssymb,amsfonts}
\usepackage[font=small,labelfont=bf]{caption}
\usepackage{xcolor}
\usepackage{cite}
\usepackage{enumerate}
\usepackage{enumitem}
\usepackage{pgf}

\usepackage{tikz}
\usetikzlibrary{arrows,automata,positioning}

\usepackage{theorem}
\newtheorem{lemma}{Lemma}
\newtheorem{remark}{Remark}

\newtheorem{assumption}{Assumption}

\newtheorem{definition}{Definition}
\newtheorem{theorem}{Theorem}
\newtheorem{problem}{Problem}
\newtheorem{example}{Example}

\newcommand{\T}{\mathcal{T}} 
\newcommand{\A}{\mathcal{A}} 
\newcommand{\B}{\mathcal{B}} 

\renewcommand{\P}{\mathcal{P}} 
\newcommand{\init}{\mathit{init}}

\newcommand{\currs}{\mathfrak{s}}
\newcommand{\currq}{\mathfrak{q}}

\newcommand{\sync}{\mathit{sync}}
\newcommand{\nosync}{\mathit{nosync}}
\newcommand{\Sync}{\mathit{Sync}}
\newcommand{\TS}{{\T=(S,s_{\init},{A} ,T)}}
\newcommand{\TSi}{{\T_i=(S_i,s_{\init,i},{A_i} ,T_i)}}
\newcommand{\N}{\mathcal{N}}

\newcommand{\model}{\mathcal{M}}
\newcommand{\I}{\mathcal{I}}
\newcommand{\D}{{D}}

\newcommand{\AP}{\Pi} 
\newcommand{\APs}{\mathbf{\Pi}}
\newcommand{\Lang}{\mathcal{L}} 
\newcommand{\Set}{\mathsf{S}} 
\newcommand{\Spec}{\mathbf{\Phi}}
\newcommand{\Epsilon}{\mathcal{E}}
\renewcommand{\i}{\iota}
\newcommand{\Nat}{\mathbb{N}} 
\newcommand{\Real}{\mathbb{R}}
\newcommand{\Next}{\mathsf{X}}
\newcommand{\Until}{\mathsf{U}}
\newcommand{\Always}{\mathsf{G}}
\newcommand{\Event}{\mathsf{F}}

\renewcommand{\epsilon}{\varepsilon}
\renewcommand{\prop}{\pi}
\newcommand{\ie}{{i.e., }}
\newcommand{\eg}{{e.g., }}

\newcommand{\h}{h}
\renewcommand{\H}{H}

\newcommand{\Alpha}{\mathbf{\Sigma}}
\renewcommand{\mod}{\mathrm{\, mod \, }}

\newcommand{\dist}{\mathrm{dist}}

\newcommand{\jana}[1]{{#1}}

\begin{document}

\begin{frontmatter}
\runtitle{Event-Based Multi-Agent Planning under Local LTL Specifications}  

\title{Multi-Agent Planning under Local LTL Specifications and Event-Based Synchronization \thanksref{footnoteinfo}} 

\thanks[footnoteinfo]{This paper was not presented at any IFAC 
meeting. Corresponding author is J. T{u}mova. 
}

\author[kth]{Jana Tumova}\ead{tumova@kth.se},    
\author[kth]{Dimos V. Dimarogonas}\ead{dimos@kth.se} 

\address[kth]{School of Electrical Engineering, KTH Royal Institute of Technology, SE-100 44 Stockholm}                                           
          
\begin{keyword} 
Temporal logic, finite state machines, formal verification, path planning, synchronization, decentralized control, robot control. 
\end{keyword}

\begin{abstract}
We study the problem of plan synthesis for multi-agent systems, to achieve complex, high-level, long-term goals that are assigned to each agent individually. As the agents might not be capable of satisfying their respective goals by themselves, requests for other agents' collaborations are a part of the task descriptions. We consider that each agent is modeled as a discrete state-transition system and its task specification takes a form of a linear temporal logic formula. 
A traditional automata-based approach to multi-agent plan synthesis from such specifications builds on centralized team planning and full team synchronization after each agents' discrete step, and thus suffers from extreme computational demands. We aim at reducing the computational complexity by decomposing the plan synthesis problem into finite horizon planning problems that are solved iteratively, upon the run of the agents. 
We introduce an event-based synchronization that allows our  approach to efficiently adapt to different time durations of different agents' discrete steps. We discuss the correctness of the solution and find assumptions, under which the proposed iterative algorithm leads to provable eventual satisfaction of the desired specifications.

\end{abstract}

\end{frontmatter}

\section{Introduction}
\renewcommand{\alph}{\Sigma}
In recent years, a considerable amount of attention has been devoted to synthesis of robot controllers for complex, high-level missions, such as ``periodically survey regions $A$, $B$, $C$, in this order, while avoiding region~$D$'', specified as temporal logic formulas. 
Many of the suggested solutions to variants of this problem rely on a hierarchical procedure \cite{hadas09TL, marius-tac2008, nok-hscc2010, kavraki-ram}: First, the dynamics of the robotic system is abstracted into a finite transition system using e.g., sampling or cell decomposition methods. Second, leveraging ideas from formal verification, a discrete plan that meets the mission is synthesized. Third, the plan is translated into a controller for the original system.
In this work, we focus on a multi-agent version of the above problem. We consider a {heterogeneous} team of robots, that are assigned a temporal logic mission each. 
As the robots may not be able to accomplish their missions without the help of the others, the specifications may contain requirements on the other team members' behavior. For instance, consider a warehouse solution with two mobile robots. A part of the first robot's mission is to load an object in region~$A$, but it is not able to load it by itself. Therefore, the mission also includes a task for the second robot, to help loading. The goal of this paper is to efficiently synthesize a plan for each agent, such that each agent's mission is met. We follow the hierarchical approach to robot controller synthesis as outlined above and we narrow our attention to the second step of the approach, i.e., to generating discrete plans. The application of the algorithm that we propose is, however, not restricted to discrete systems: For the first step of the hierarchical approach, numerous methods for discrete modeling of robotic systems can be used (see,~\eg\cite{hadas09TL, marius-tac2008, nok-hscc2010, lavalle} and the references therein); for the third step, low-level controllers exist that can drive a robot from any position within a region to a goal region (see, \eg\cite{Belta-TAC06}). {The agents can, but do not have to, mutually synchronize after the execution of their respective discrete steps. The desired plans thus comprise not only of the agents' discrete steps to be taken, but also their synchronizations. Besides the satisfaction of all agents' missions, our goal is to avoid unnecessary synchronization in order to improve the team performance.}
 
As a mission specification language, we use Linear Temporal Logic (LTL), for its resemblance to natural language \cite{hadas-icra2012}, and expressive power. 
Here, we built LTL formulas over services, i.e., events of interest associated with execution of certain actions rather than over atomic propositions, i.e., inherent properties of the system states. 
Instead of evaluation of the specification as a conjunction of LTL formulas over the whole team behaviors, we propose the notion of satisfaction of an LTL formula from local perspective. This way, the problem of finding a collective team behavior is decomposed into several subproblems, enabling us to avoid the straightforward, but expensive fully centralized planning.
The contribution of this paper can be summarized as the introduction of an efficient, iterative, finite horizon planning technique in the context of bottom-up plan synthesis for multi-agent systems from local LTL specifications. To our best knowledge, such an approach has not been taken to address the multi-agent LTL planning 
before. 
Our algorithm is adaptive in the sense that even if the real behavior of the team is not as planned due to unpredictable time durations of the agents' steps, the event-based synchronization and replanning still guarantees the satisfaction of all the tasks. This feature can be especially beneficial in heterogeneous multi-robot motion and task planning problems, where individual robots traverse their common environment at different speeds. This paper builds on our earlier work in~\cite{acc14}. In addition, it relaxes the assumption that the agents synchronize after every  discrete step of theirs and introduces the event-based synchronization and replanning.

Multi-agent planning from temporal logic specification has been explored in several recent works. Planning from computational tree logic was considered in~\cite{quo-icra2004}, whereas in~\cite{loizou-cdc2005,marius-cdc2011}, the authors focus on planning behavior of a team of robots from a single, global LTL specification. A fragment of LTL has been considered as a specification language for vehicle routing problems in~\cite{sertac-ijnc2010}, and a general reactivity LTL fragment has been used in~\cite{lygeros-ecc2013}. 
Decentralized control of a robotic team from local LTL specification with communication constraints is proposed in~\cite{dimos-cdc12}. However, the specifications there are truly local and the agents do not impose any requirements on the other agents' behavior. Thus, the focus of the paper is significantly different to ours.
As opposed to our approach, in~\cite{yushan-tr2012,alphan-ijrr2013}, a top-down approach to LTL planning is considered; the team is given a global specification and an effort is made to decompose the formula into independent local specifications that can be treated separately for each robot.
In~\cite{meng-ijrr2015}, bottom-up planning from LTL specifications is considered, and a partially decentralized solution is proposed that takes into account only clusters of dependent agents instead of the whole group. A huge challenge of the previous approach is its extreme computational complexity, which we tackle in this work by
applying receding horizon approach to multi-agent planning. 
Receding horizon approach was leveraged also in \cite{nok-hscc2010} to cope with uncertain elements in an environment in single-robot motion planning. To guarantee the satisfaction of the formula, we use an attraction-type function that guides the individual agents towards a progress within a finite planning horizon; similar ideas were used in~\cite{dennis-rh2,maja} for a single-agent LTL planning to achieve a locally optimal behavior.

The rest of the paper is structured as follows. In Sec.~\ref{sec:prelims}, we fix the preliminaries. Sec.~\ref{sec:pf} introduces the problem and summarizes our approach. In Sec.~\ref{sec:solution}, the details of the solution are provided. In Sec.~\ref{sec:discussion}, we provide analysis and discussion of the solution. We present simulation results in Sec.~\ref{sec:simulations}, and we conclude in Sec.~\ref{sec:summary}.

\section{Preliminaries}
\label{sec:prelims}

Given a set $\Set$, let $2^\Set$, and $\Set^\omega$ denote the set of all subsets of $\Set$, and the set of all infinite sequences of elements of $\Set$, respectively. A finite \jana{or} infinite sequence of elements of $\Set$ is called a finite \jana{or} infinite word over $\Set$, respectively. The $i$-th element of a word $w$ is denoted by $w(i)$. A {subsequence} of  an infinite word $w = w(1)w(2)\ldots$ is a finite or infinite sequence of its elements $w(i_1)w(i_2)\ldots$, where $\forall 1 \leq j. \ 1 \leq i_j \leq i_{j+1}$. A {factor} of  $w$ is a continuous, finite or infinite, subsequence $w(i)w(i+1)\ldots$, where $1 \leq i$. A {prefix} of $w$ is a finite factor starting at $w(1)$, and a {suffix} of $w$ is an infinite factor.
{$\Nat$ and $\Real_0^+$ denote positive integers and non-negative real numbers, respectively.}

A \emph{transition system (TS)} is a tuple $\TS$, where
$S$ is a finite set of states;
$s_{init} \in S$ is the initial state;
$A$ is a finite set of actions; and
{$T \subseteq S \times A \rightarrow S$} is a partial deterministic transition function.
For simplicity, we denote a transition $T(s,\alpha) = s'$ by $s \xrightarrow{\alpha} s'$. A \emph{trace} of $\T$ is an infinite alternating sequence of states and actions $\tau = s_1\alpha_1s_2\alpha_2\ldots$, such that $s_1=s_\init$, and for all $i \geq 1$, $s_i \xrightarrow{\alpha_i} s_{i+1}$. A \emph{trace fragment} $\hat \tau$ 
is a finite factor of a trace $\tau$ that begins and ends with a state.

A \emph{linear temporal logic (LTL) formula} $\phi$ over {the set of atomic propositions $\AP$} is defined inductively: (i) $\prop \in \AP$ is a formula, and (ii) if $\phi_1$ and $\phi_2$ are formulas, then $\phi_1 \lor \phi_2$, $\lnot \phi_1$, $\Next\, \phi_1$, $\phi_1\,\Until\,\phi_2$, $\Event \, \phi_1$, and $\Always \, \phi_1$ are each {a formula}, where $\neg$ and $\vee$ are standard Boolean connectives, and $\Next$, $\Until$, $\Event$, and  $\Always$  are temporal operators.
The semantics of LTL are defined over infinite words over~$2^\AP$. $\pi \in \AP$ is satisfied on {$w = \varpi_1\varpi_2\ldots$} if $\pi \in \varpi_1$. $\Next \, \phi$ holds true if $\phi$ is satisfied on the word   that begins in the next position {$\varpi_2$}, $\phi_1 \, \Until\, \phi_2$ states that $\phi_1$ has to be true until $\phi_2$ becomes true, and $\Event \, \phi$ and $\Always \, \phi$ are true if $\phi$ holds on $w$ eventually, and always, respectively. \jana{We denote the satisfaction of $\phi$ on a word $w$ as $w \models \phi$.} 
The set of all words accepted by an LTL formula $\phi$ is  $\Lang(\phi)$. For full details see, e.g.,~\cite{principles}.

An \emph{automaton} is a tuple $\A =  (Q,q_{init},\Sigma,\delta,F)$, where
$Q$ is a finite set of states; 
$q_{init}\in Q$ is the initial state; 
$\Sigma$ is an input alphabet; 
$\delta \subseteq Q \times \Sigma \times Q$ is a non-deterministic transition relation; and
$F$ is an accepting condition.
It is \emph{deadlock-free} if $\forall q \in Q, \sigma \in \Sigma. \ \delta(q,\sigma) \neq \emptyset$.
We define the set of states $\hat \delta^k(q)$ reachable from $q \in Q$ in exactly $k$ steps inductively as
$ \hat \delta^0(q)   = \{q\}, \text{ and} $ 
$ \hat \delta^{k}(q)  = \bigcup_{q' \in \hat \delta^{k-1}(q)} \{q'' \mid \exists \, \sigma \in \Sigma. \, (q',\sigma,q'') \in \delta\}, \forall k \geq 1$.
A \emph{B\"uchi automaton (BA)} is an automaton with the accepting condition $F \subseteq Q$.
A \emph{run} of the BA $\B$ from $q_1 \in Q$ \emph{over} {$w=\sigma_1\sigma_2\ldots \in \Sigma^\omega$}  is a sequence $\rho=q_1q_2\ldots$, such that $\forall i\geq 1.\ (q_i,\sigma_i,q_{i+1}) \in \delta$.
A run $\rho$ is \emph{accepting} if it intersects $F$ infinitely many times. A word $w$ is \emph{accepted} by $\B$ if there exists an accepting run over~$w$ from the state $q_\init$. The set of all words accepted by $\B$ is $\Lang(\B)$. Any automaton $(Q,q_\init,\Sigma,\delta,F)$ can be viewed as a graph $(V,E)$ with the vertexes $V= Q$ and the edges~$E$ given by $\delta$ in the expected way. The standard notation then applies: A \emph{path} is a finite alternating sequence of states and transition labels $q_i{\sigma_i}q_{i+1}\ldots q_{k-1}{\sigma_{k-1}}q_k$, such that $\forall i \leq j< k. \ (q_j,\sigma_j, q_{j+1})\in \delta$. 
$\dist(q,q')$ denotes the length of the \emph{shortest path} between $q$ and $q'$, \ie the minimal number of states in  a path $q\ldots q'$. If no such path exists, then $\dist(q,q')=\infty$. If $q = q'$, then $\dist(q,q')=0$. The shortest path can be computed using Dijkstra algorithm~(see, e.g., \cite{cormen}).

\section{Problem Formulation and Approach}
\label{sec:pf}

Two general viewpoints can be taken in multi-agent planning: either the system acts as a team with a common goal, or the agents have their own, independent tasks. Although we permit each agent's task to involve requirements on the others, we adopt the second viewpoint; to decide whether the agents' tasks are met, we do not look at the global team behavior. In contrast, we propose the concept of local specification satisfaction to determine whether an agent's task is fulfilled from its own perspective. 
Our goal is to synthesize a plan for each agent that comprises of (i)~the actions to be executed and (ii)~the synchronization requests imposed on the other agents. Together, the strategies have to ensure the local satisfaction of each of the task specifications regardless of the time duration of the planned action executions. 

We consider $N$ agents (e.g., robots in a partitioned environment). Each agent $i \in \N$, where $\N = \{1,\ldots, N\}$ has action execution and synchronization capabilities.

\textbf{Action Execution Capabilities.} The agent $i$'s action execution capabilities are modeled as a   TS $\TSi$, whose states correspond to states of the agents (\eg~the locations of the robots in the regions of their environment). 
The actions $A_i$ represent abstractions of the agent's low-level controllers, and the transitions $T_i$ correspond to the agent's capabilities to execute the actions (\eg the ability of the robots to move between two regions). The traces are, roughly speaking, the abstractions of the agents' long-term behaviors (\eg the robots' trajectories). For the simplicity of  the presentation, we assume that each state $s \in S_i$ is reachable from all states $s'\in S_i$ via a sequence of transitions. 
Each of the agents' action executions takes a certain amount of time. Given a trace $\tau = s_1\alpha_1s_2\alpha_2\ldots$ of $\T_i$, 
we denote by $\Delta_{\alpha_j} \in \Real_0^+$ the time \emph{duration} of the transition $s_j\xrightarrow{\alpha_j}s_{j+1}$.
Note that a transition duration is arbitrary and unknown prior its execution, and that the execution of the same action $\alpha$ may induce different transition durations in its different instances on a trace. 
{The durations are not explicitly modeled in the TS $\T_i$ due to the fact that they are unknown, but, as we will discuss later on, their history plays an important role in defining the semantics of the agent's behaviors, their interactions and satisfaction of their tasks.}

\textbf{Synchronization Capabilities.} The agents have also the ability to \emph{synchronize}, {i.e.,} to wait for each other and to proceed with the further  execution simultaneously. {Through the simultaneous execution of certain transitions, the agents have the ability to collaborate (e.g., an agent loading heavy goods may need a second agent simultaneously helping to load it). 
The synchronization is modeled through the \emph{synchronization requests}, but the particular implementation of the synchronization scheme is beyond the scope of this paper. While being in a state $s$, an agent $i$ can send a request $sync_i$ to the set of agents $\N$ notifying them that it is ready to synchronize. Then, before proceeding with the execution of any action $\alpha \in A_i$, it has to wait till $sync_{i'}$ has been sent by each agent $i' \in \N$, i.e., till the moment when each agent $i' \in \N$ is ready to synchronize, too. The synchronization is immediate once each of the agents $i' \in \N$ has sent its request  $sync_{i'}$ and all agents in $\N$ start executing the next action at the same time. Alternatively, an agent $i$ indicates that it does not need to synchronize through requesting $\nosync_i$. The set of all synchronization requests of an agent $i$ is thus $\Sync_i = \{\sync_i, \nosync_i\}$.
For simplicity, we assume that each agent sends a synchronization request instantly once it completes an action execution and that it starts executing an action instantly once it synchronizes with the other agents. In other words, no time is spent idling in our system model. 
This does not prevent us to capture the agents' abilities of staying in their respective states. Instead of idling, we include a so-called self-loop $s \xrightarrow{\alpha} s$ for some $\alpha \in A_i$, all $s \in S_i$, and all $\T_i$,  $i \in \N$.
Given a trace $\tau_i = s_{i,1}\alpha_{i,1}s_{i,2}\alpha_{i,2}\ldots$ of $\T_i$,  we denote by $\Delta_{s_{i,j}} \in \Real_0^+$ the time duration of the synchronization requested in the state $s_{i,j}$. 
Note that if the request $\nosync_i$ has been sent in $s_{i,j}$, then $\Delta_{s_{i,j}} = 0$.

\begin{remark}
In order to accomodate synchronization with a subset of agents, 
we can  parametrize the synchronization requests $\Sync_i = \{\sync_i(M) \mid \{i\} \subseteq M \subseteq \N\}$. The use of such a definition is discussed later in Remark~\ref{remark:sync2}.
\label{remark:sync1}
\end{remark}

\textbf{Services.} Each of the agents' tasks is given in terms of temporal requirements on events of interest, which we call \emph{services}. The set of all services that can be \emph{provided} by an agent $i\in \N$ is $\AP_i$. Services are {provided} within the agents' transitions; each action $\alpha \in A_i$ is \emph{labeled} either with 
(i) a service set $\varpi \in 2^{\AP_i}$ provided upon the execution of $\alpha$, or 
(ii) a special \emph{silent} service set $\Epsilon_i = \{\epsilon_i\}$, $\epsilon_i \not \in {\AP_i}$ indicating that $\alpha$ is not associated with any event of interest.
{Hence, two additional components of the agent $i$'s model are
the set of all available services $\AP_i$ and the action-labeling function $L_i: A_i \to 2^{\Pi_i} \cup 2^{\Epsilon_i}$.} Note that we specifically distinguish between a silent service set $\Epsilon_i$ and an empty service set $\{\}$. 
The self-loops of type $s \xrightarrow{\alpha} s$ introduced above to model staying are naturally labeled with $\Epsilon_i$.
Without loss of generality, we assume that $\AP_i \cap \AP_{i'} = \emptyset$, for all $i,i' \in \N$, $i\neq i'$. 
{As it will become clear later, the choice of non-silent and silent services in place of the traditional atomic propositions  is motivated by the nature of multi-agent planning, where the agents are concerned about each other only at selected times, e.g., when collaboration is required.}

Finally, we model an agent $i \in \N$ as  the tuple 
$\model_i = (\T_i, \Sync_i, \AP_i, L_i).$

\begin{example}
An illustrative example of three mobile robots in a common partitioned workspace is depicted in Fig.~\ref{fig:example1}. The agents can transit between the adjacent cells that are not separated by a wall
or stay where they are. The former transitions are labeled with $\Epsilon_i$, while non-silent services are associated with some of the latter ones. Namely, agent 1 can load ($l_H,l_A,l_B$), carry, and unload ($u_H,u_A,u_B$) a heavy object $H$ or a light object $A$, $B$, {in the green cells}. Agent 2 can help to load object $H$ ($h_H$), and execute simple tasks in the purple regions ($t_1-t_5$). Agent 3 is capable of taking a snapshot of the rooms ($s_{1}-s_{5}$) when being present in there. 
\label{example: running}
\begin{figure}[!t]
\begin{center}
\includegraphics[width=0.7\linewidth]{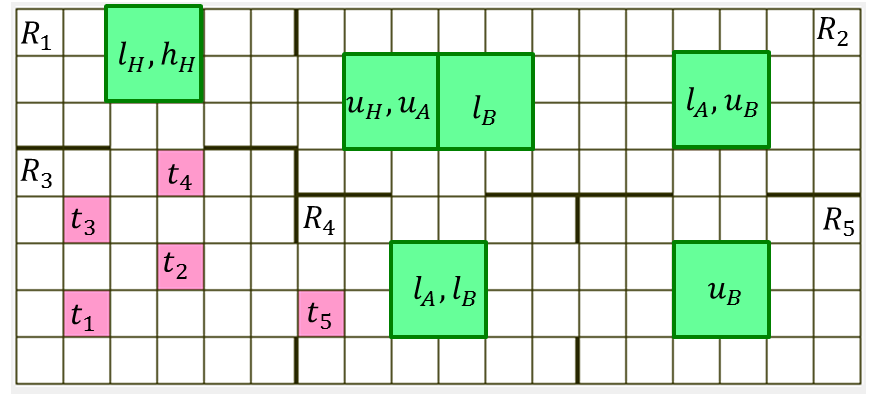}
\caption{\scriptsize An example of an environment partitioned into cells with rooms $R_1,\ldots, R_5$. Loading/unloading points are in green (light). Simple tasks $t_1,\ldots,t_5$ can be executed in purple (dark) regions.}
\label{fig:example1}
\end{center}
\end{figure}
\end{example}

\textbf{Behaviors.} The \emph{behavior} of an agent $i$ is defined by the actions it executes, its synchronizations with the other agents, and the time instants when the action executions and the synchronizations take place.
\begin{definition}[Behavior] A behavior of an agent $i$ is a tuple \jana{$\beta_i = (\tau_i,\gamma_i, \mathbb T_i)$}, where
\jana{$\tau_i = s_{i,1}\alpha_{i,1}s_{i,2}\alpha_{i,2}\ldots$} is a trace of $\T_i$;
\jana{$\gamma_i = r_{i,1}r_{i,2}\ldots $} is a \emph{synchronization sequence}, where \jana{$r_{i,j} \in {\Sync_i}$} is the synchronization request sent at the state \jana{$s_{i,j}$}; and
$\mathbb{T}_i = t_{s_{i,1}}t_{\alpha_{i,1}}t_{s_{i,2}}t_{\alpha_{i,2}}\ldots$ is a non-decreasing \emph{behavior time sequence}, where  
$t_{s_{i,j}}$ is the time instant when the synchronization request $r_{i,j}$ was sent, and $t_{\alpha_{i,j}}$ is the time instant when the action $\alpha_{i,j}$ started being executed. The following hold: $t_{s_{i,1}}=0$, and for all $j \geq 1$, $t_{s_{i,j+1}}-t_{\alpha_{i,j}} = \Delta_{\alpha_{i,j}}$, and $t_{\alpha_{i,j}} - t_{s_{i,j}}=\Delta_{s_{i,j}}$.
\label{def:behavior}
\end{definition}

The notion of behavior, however, does not reflect the above described synchronization scheme. To that end, we define \emph{compatible behaviors}. 
In what follows, the behavior of $i \in \N$ is denoted by $\beta_i = (\tau_i,\gamma_i,\mathbb T_i)$, where $\tau_i = s_{i,1}\alpha_{i,1}s_{i,2}\alpha_{i,2}\ldots$, $\gamma_i = r_{i,1}r_{i,2}\ldots $, and $\mathbb{T}_i = t_{s_{i,1}}t_{\alpha_{i,1}}t_{s_{i,2}}t_{\alpha_{i,2}}\ldots$.

\begin{definition}[Compatible behaviors]
A set of behaviors of the agents in $\N$ is \emph{compatible}, if the following holds for all $i \in \N$, and $j \geq 1$: Suppose that $r_{i,j} = sync_i$. Then for all $i' \in \N$, there exists a \emph{matching index} $j' \geq 1$, such that $r_{i',j'} = sync_{i'}$, and $t_{\alpha_{i,j}} = t_{\alpha_{i',j'}}$. Furthermore, there exists $i' \in \N$, such that $t_{s_{i',j'}} = t_{\alpha_{i',j'}}$, i.e., such that $\Delta_{s_{i',j'}} = 0$, for the matching index $j'$.
\label{def:compatible}
\end{definition}

Suppose that the trace $\tau_i$, the synchronization sequence $\gamma_i$, and the transition time durations $\Delta_{\alpha_{i,1}},\Delta_{\alpha_{i,2}},\ldots$ are all fixed for all $i \in \N$. Then there exists at most one collection of synchronization time durations $\Delta_{s_{i,1}}\Delta_{s_{i,2}}\ldots$, $i \in \N$ that yields a set of compatible behaviors $\{\beta_i = (\tau_i, \gamma_i, \mathbb T_i) \mid i \in \N\}$. In other words, if the agents follow the synchronization scheme, the existence and the values of their synchronization time durations and hence also their behavior time sequences are uniquely determined by $\tau_i$, $\gamma_i$, and $\Delta_{\alpha_{i,1}},\Delta_{\alpha_{i,2}},\ldots$ given for all $i \in \N$.

\textbf{Specifications.} The individual agents' tasks  may concern the respective agent's services as well as the services of the others. Formally, each of the agents is given an LTL formula $\phi_i$ over 
$\APs_i = \bigcup_{i' \in \D_i} \AP_{i'}$, for some {$\{i\} \subseteq \D_i \subseteq \N$}. Loosely speaking, the satisfaction of an agent's formula depends on, and only on the behavior of the subset of agents {$\D_i$}, including the agent itself. 

\addtocounter{example}{-1}
\begin{example}[continued]
The robots are assigned the following collaborative tasks. Agent 1 needs the help of agent 2 with loading the heavy object. Then, it should carry the heavy object to an unloading point and unload it. Then, it should periodically load and unload both light objects ($ \phi_1 = \Event (l_H \wedge h_H \wedge \Next \, u_H \wedge \bigwedge_{i \in \{A,B\}} \Always \Event \, (l_i \wedge \Next u_i))$). Agent~2 should periodically execute the simple tasks $t_1,\ldots,t_5$, in this order. Furthermore, it requests agent 3 to witness the execution $t_5$, by taking a snapshot of the room $R_4$ at the moment of the execution ($ \phi_2 = \Always \Event \ (t_1 \wedge \Next \ (t_2 \wedge \Next \ (t_3 \wedge \Next \ (t_4 \wedge \Next \ t_5 \wedge s_{4})))))$). Agent 3 should patrol rooms $R_2,R_4,R_5$ ($\phi_3 = \bigwedge_{i\in \{2,4,5\}} \Always \Event \, s_{i}$).
\label{example:running2}
\end{example}

Let us now introduce the notation needed for formalizing the specification satisfaction.
Consider for a moment a single agent $\model_i = (\T_i,\Sync_i, \AP_i, L_i)$, and its behavior $\beta_i$, where, for simplicity of notation in the following paragraphs, we use $\beta_i = (\tau, \gamma, \mathbb T)$, where $\tau = s_{1}\alpha_{1}s_{2}\alpha_{2}\ldots$, and  $\mathbb{T} = t_{s_1}t_{\alpha_1}t_{s_2}t_{\alpha_2}\ldots$.
We denote by  $v_\tau = \varpi_1\varpi_2\ldots = L_i(\alpha_{1})L_i(\alpha_{2})\ldots  \in (2^{\Pi_i} \cup 2^{\Epsilon_i})^\omega$ the unique \emph{service set sequence} associated with $\tau$. 
The \emph{word} $w_\tau$ produced by $\tau$ is the subsequence of the non-silent elements of this sequence; $w_\tau = \varpi_{\i_1}\varpi_{\i_2}\ldots \in (2^{\Pi_i})^\omega$, such that $\varpi_1, \ldots, \varpi_{\i_1-1} = \Epsilon_i$, and for all $j \geq 1$, $\varpi_{\i_j} \neq \Epsilon_i$ and $\varpi_{\i_j+1}, \ldots, \varpi_{\i_{j+1}-1} = \Epsilon_i$.
With a slight abuse of notation, we use $\mathbb{T}(\tau) = t_1t_2\ldots = t_{\alpha_1}t_{\alpha_2}\ldots$ to denote the \emph{trace time sequence}, i.e., the subsequence of $\mathbb T$ when the (both silent and non-silent) services are provided.
Furthermore, $\mathbb{T} (w_\tau)= t_{\i_1}t_{\i_2}\ldots$ denotes the \emph{word time sequence}, i.e., the subsequence of $\mathbb{T}(\tau)$ that corresponds to the times when the non-silent services are provided. 
The word {$w_\tau$} and the word time sequence {$\mathbb{T}(w_\tau)$} might be finite as well as infinite. Since in this work we are interested in infinite, recurrent behaviors, we will consider as \emph{valid} traces only those that produce infinite words. 
Consider a trace $\tau$ of $\T_i$ with the service set sequence $v_\tau = \varpi_1\varpi_2\ldots$, and the trace time sequence $\mathbb{T}(\tau) = t_{1}t_{2}\ldots$. 
The service set $v_\tau(t) \in 2^{\Pi_i} \cup 2^{\Epsilon_i}$ provided at time $t \in \Real_0^+$ is
$v_\tau(t) = \varpi_j  \text{ if } t=t_j \text{ for some } j \geq 1$, and $ \Epsilon_i  \text{ otherwise.} $
The satisfaction of each LTL formula $\phi_i$ is interpreted locally, from the agent $i$'s point of view, based on the word {$w_{\tau_i}$} it produces and on the services of agents $i' \in \D_i$ provided at the time instances $\mathbb T(w_{\tau_i})$ when $i$ provides a non-silent service set itself.


\begin{definition}[Local LTL satisfaction]
Let $\tau$ be a trace of $\T_i$ with the word time sequence $\mathbb T(w_{\tau}) = t_{\i_1}t_{\i_2}\ldots$.  The \emph{word} produced by a set of compatible behaviors  $\mathfrak{B}_i =  \{\beta_{i'} \mid i' \in \D_i\}$ is $w_{\mathfrak{B}_i} = \omega_{\i_1}\omega_{\i_2}\ldots, \text{ where} $
$\omega_{\i_j} = \big(\bigcup_{i' \in \D_i} v_{\tau_{i'}}(t_{\i_j})\big) \cap  {\APs_i}, \text{ for all } j\geq 1.$
The set of behaviors $\mathfrak{B}_i$ is \emph{valid}  if $w_{\mathfrak{B}_i}$ is infinite.
The set of compatible behaviors $\mathfrak{B}_i$ \emph{locally satisfies} $\phi_i$ for the agent~$i$, $\mathfrak{B}_i \models \phi_i$, iff $\mathfrak{B}_i$ is valid and $w_{\mathfrak{B}_i} \models \phi_i$.
\label{def:localsat}
\end{definition}

Note that even if $\mathfrak{B} = \mathfrak{B}_i = \mathfrak{B}_j$, it may be the case that $w_{\mathfrak B_i} \neq w_{\mathfrak B_j}$ and  $\mathfrak{B_i} \models \phi$, but $\mathfrak{B_j} \not \models \phi$.
Let us provide some intuitive insight into the local LTL satisfaction through the following example.

\begin{example}
Consider two agents represented by models $\model_1 = (\T_1, \Sync_1, \AP_1, L_1)$, and $\model_2 = (\T_2, \Sync_2, \AP_2, L_2)$, with $\AP_1 = \{a\}$, $\AP_2 = \{b\}$, and their LTL specifications $\phi_1 = a \, \wedge \, \Next \,  (a \, \wedge \, b)$,  $\phi_2 = b \, \wedge \, \Next \, (b \, \wedge \, a)$, respectively. Note that both $1 \in \D_2$ and $2 \in \D_1$, and hence $\APs_1 =\APs_2 = \{a,b\}$. 
Let $\beta_i = (\tau_i,\gamma_i,\mathbb T_i)$ be the respective behavior of the agent $i \in \{1,2\}$ with the trace $\tau_i$ as illustrated in Fig.~\ref{fig:exampleA}, $\gamma_i = r_{i,1}r_{i,2}\ldots$, where $r_{i,j} = \nosync_i$, and $\mathbb T_i =  t_{s_{i,1}}t_{\alpha_{i,1}}t_{s_{i,2}}t_{\alpha_{i,2}}\ldots$, where $ t_{s_{i,j}} = t_{\alpha_{i,j}} = j$, for all $j \geq 1$. Fig.~\ref{fig:exampleA} depicts the service set sequence, the trace time sequence, the word and the word time sequence for both agents. Specifically, $\mathbb{T}(w_{\tau_1}) = t_1  t_5 t_6 \ldots = 0 \ 4 \ 5 \ldots$ and hence, the word produced by
$\mathfrak B_1$ is $w_{\mathfrak B_1} = \omega_1  \omega_5 \omega_6 \ldots$, 
where $\omega_1 = (v_{\tau_1}(0) \, \cup \, v_{\tau_2}(0))  \, \cap \, {\APs_1} = (\{a\} \cup \, \Epsilon_2) \, \cap \, {\APs_1} = \{a\}$, $\omega_5 =  (v_{\tau_1}(4) \, \cup \, v_{\tau_2}(4))  \, \cap \, {\APs_1} = \{a,b\} \, \cap \, {\APs_1} = \{a,b\}$, and $\omega_6 =  (v_{\tau_1}(5) \, \cup \, v_{\tau_2}(5))  \, \cap \, {\APs_1} =  \{\} \, \cap \, {\APs_1} = \{\}$. Since $w_{\mathfrak B_1}$ is valid and satisfies $\phi_1$, we conclude that $\mathfrak B_1$ locally satisfies $\phi_1$. In contrast, \jana{$\mathbb{T}(w_{\tau_2})$} = $t_2  t_3 t_5 t_6 \ldots = 1 \ 2 \ 4 \ 5 \ldots$, and
 the word produced by $\mathfrak{B}_2$ is $w_{\mathfrak{B}_2} = \omega_2 \omega_3 \omega_5 \omega_6 \ldots$, where $\omega_2 = \{b\}$, $\omega_3 = \{b\}$, $\omega_5 = \{a,b\}$, and $\omega_6 = \{\}$. Although $\mathfrak{B}_2$ is valid, it does not satisfy formula $\phi_2$. 
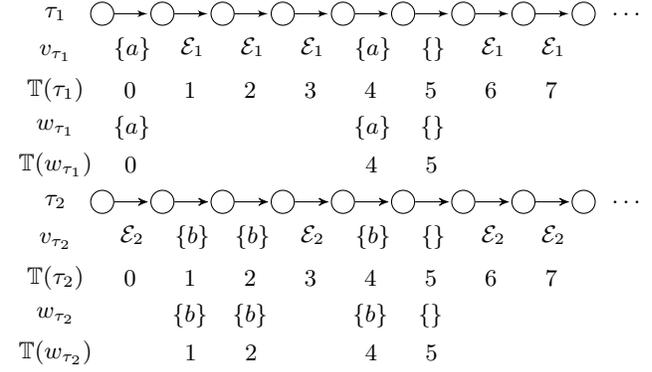
\begin{figure}[!h]
\footnotesize
\begin{center}
\usetikzlibrary{arrows,positioning,automata,shadows,fit,shapes,positioning}
\begin{tikzpicture}[->,>=latex',shorten >=1pt,auto, node distance=0.8cm,initial text=]
  \tikzstyle{every state}=[draw]
	  \node[] (s0) {$\tau_1$}; 
           \node[] (s00)  [below = 0.1 of s0] {$v_{\tau_1}$}; 
	  \node[state, minimum size=1ex] (s1) [right = 0.2cm of s0] {}; 
	  \node[state, minimum size=1ex] (s2) [right of = s1] {}; 
	  \node[state, minimum size=1ex] (s3) [right of = s2] {}; 
	  \node[state, minimum size=1ex] (s4) [right  of = s3] {}; 
	  \node[state, minimum size=1ex] (s5) [right  of = s4] {}; 
	  \node[state, minimum size=1ex] (s6) [right  of = s5] {}; 
	  \node[state, minimum size=1ex] (s7) [right  of = s6] {}; 
	  \node[state, minimum size=1ex] (s8) [right  of = s7] {}; 
	  \node[state, minimum size=1ex] (s9) [right  of = s8] {};
	  \node[] (s10) [right = 0.1cm of s9] {$\ldots$};  

          \node[] (s01) [below = 0cm of s00]{$\mathbb T(\tau_1)$}; 
	  \node[] (s11) [right = 0.3cm of s01] {0}; 
	  \node[] (s21) [right of = s11] {1}; 
	  \node[] (s31) [right of = s21] {2}; 
	  \node[] (s41) [right  of = s31] {3}; 
	  \node[] (s51) [right  of = s41] {4}; 
	  \node[] (s61) [right  of = s51] {5}; 
	  \node[] (s71) [right  of = s61] {6}; 
	  \node[] (s81) [right  of = s71] {7}; 

          \node[] (s011) [below = 0cm of s01]{$w_{\tau_1}$}; 
	  \node[] (s111) [right = 0.28cm of s011] {$\{a\}$}; 
	  \node[] (s211) [right of = s111] {}; 
	  \node[] (s311) [right of = s211] {}; 
	  \node[] (s411) [right  of = s311] {}; 
	  \node[] (s511) [right  of = s411] {$\{a\}$}; 
	  \node[] (s611) [right  of = s511] {$\{\}$}; 
	  \node[] (s711) [right  of = s611] {}; 
	  \node[] (s811) [right  of = s711] {};

          \node[] (s0111) [below = 0cm of s011]{$\mathbb T(w_{\tau_1})$}; 
	  \node[] (s1111) [right = 0.2cm of s0111] {$0$}; 
	  \node[] (s2111) [right of = s1111] {}; 
	  \node[] (s3111) [right of = s2111] {}; 
	  \node[] (s4111) [right  of = s3111] {}; 
	  \node[] (s5111) [right  of = s4111] {$4$}; 
	  \node[] (s6111) [right  of = s5111] {$5$}; 
	  \node[] (s7111) [right  of = s6111] {}; 
	  \node[] (s8111) [right  of = s7111] {};

  \path (s1) edge node [below = 0.2cm] {$\{a\}$} (s2);
  \path (s2) edge node [below = 0.2cm] {$\Epsilon_1$} (s3);
  \path (s3) edge node [below = 0.2cm] {$\Epsilon_1$} (s4);
  \path (s4) edge node [below = 0.2cm] {$\Epsilon_1$} (s5);
  \path (s5) edge node [below = 0.2cm] {$\{a\}$} (s6);
  \path (s6) edge node [below = 0.2cm] {$\{\}$} (s7);
  \path (s7) edge node [below = 0.2cm] {$\Epsilon_1$} (s8);
  \path (s8) edge node [below = 0.2cm] {$\Epsilon_1$} (s9);
\end{tikzpicture}
\smallskip
\begin{tikzpicture}[->,>=stealth',shorten >=1pt,auto, node distance=0.8cm,initial text=]
  \tikzstyle{every state}=[draw]
	  \node[] (s0) {$\tau_2$}; 
      \node[] (s00)  [below = 0.1cm of s0] {$v_{\tau_2}$}; 
	  \node[state, minimum size=1ex] (s1) [right = 0.2cm of s0] {}; 
	  \node[state, minimum size=1ex] (s2) [right of = s1] {}; 
	  \node[state, minimum size=1ex] (s3) [right of = s2] {}; 
	  \node[state, minimum size=1ex] (s4) [right  of = s3] {}; 
	  \node[state, minimum size=1ex] (s5) [right  of = s4] {}; 
	  \node[state, minimum size=1ex] (s6) [right  of = s5] {}; 
	  \node[state, minimum size=1ex] (s7) [right  of = s6] {}; 
	  \node[state, minimum size=1ex] (s8) [right  of = s7] {}; 
	  \node[state, minimum size=1ex] (s9) [right  of = s8] {};
	  \node[] (s10) [right = 0.1 cm  of s9] {$\ldots$};  

      \node[] (s01) [below = 0cm of s00]{$\mathbb T(\tau_2)$}; 
	  \node[] (s11) [right = 0.3cm of s01] {0}; 
	  \node[] (s21) [right of = s11] {1}; 
	  \node[] (s31) [right of = s21] {2}; 
	  \node[] (s41) [right  of = s31] {3}; 
	  \node[] (s51) [right  of = s41] {4}; 
	  \node[] (s61) [right  of = s51] {5}; 
	  \node[] (s71) [right  of = s61] {6}; 
	  \node[] (s81) [right  of = s71] {7}; 

      \node[] (s011) [below = 0cm of s01]{$w_{\tau_2}$}; 
	  \node[] (s111) [right = 0.5cm of s011] {}; 
	  \node[] (s211) [right of = s111] {$\{b\}$}; 
	  \node[] (s311) [right of = s211] {$\{b\}$}; 
	  \node[] (s411) [right  of = s311] {}; 
	  \node[] (s511) [right  of = s411] {$\{b\}$}; 
	  \node[] (s611) [right  of = s511] {$\{\}$}; 
	  \node[] (s711) [right  of = s611] {}; 
	  \node[] (s811) [right  of = s711] {};

          \node[] (s0111) [below = 0cm of s011]{$\mathbb T(w_{\tau_2})$}; 
	  \node[] (s1111) [right = 0.28cm of s0111] {}; 
	  \node[] (s2111) [right of = s1111] {1}; 
	  \node[] (s3111) [right of = s2111] {2}; 
	  \node[] (s4111) [right  of = s3111] {}; 
	  \node[] (s5111) [right  of = s4111] {$4$}; 
	  \node[] (s6111) [right  of = s5111] {$5$}; 
	  \node[] (s7111) [right  of = s6111] {}; 
	  \node[] (s8111) [right  of = s7111] {};

	  \path (s1) edge node [below = 0.2cm]  {$\Epsilon_2$} (s2);
	  \path (s2) edge node [below = 0.2cm]  {$\{b\}$} (s3);
	  \path (s3) edge node [below = 0.2cm] {$\{b\}$} (s4);
	  \path (s4) edge node [below = 0.2cm]  {$\Epsilon_2$} (s5);
	  \path (s5) edge node [below = 0.2cm]  {$\{b\}$} (s6);
	  \path (s6) edge node [below = 0.2cm]  {$\{\}$} (s7);
	  \path (s7) edge node [below = 0.2cm]  {$\Epsilon_2$} (s8);
	  \path (s8) edge node [below = 0.2cm]  {$\Epsilon_2$} (s9);	  
\end{tikzpicture}
\end{center}
\normalsize
\caption{The trace, the service set the trace time sequences, the word and the word time sequence of the agents $1$, and $2$.}
\label{fig:exampleA}
\end{figure}
\end{example}

\begin{problem}
\label{prob:main}
\emph{Consider} a set of agents $\N = \{1,\ldots,N\}$, and suppose that each agent $i \in \N$ is modeled as a tuple $\model_i = (\T_i, \Sync_i, \AP_i, L_i)$, 
and assigned a task in the form of an LTL formula $\phi_i$ over 
$\APs_i = \bigcup_{i' \in \D_i} \AP_{i'}$, for some $\{i\}\subseteq D_i \subseteq \N$.
\emph{For each $i \in \N$ find} a trace $\tau_i = s_{i,1}\alpha_{i,1}s_{i,2}\alpha_{i,2}\ldots$ of $\T_i$, and a synchronization sequence $\gamma_i$ over $\Sync_i$,
with the property that regardless of the values of the transition time durations $\Delta_{\alpha_{i,1}}, \Delta_{\alpha_{i,2}}\ldots \in \Real_0^+$, $i \in \N$, the set of the agents' behaviors $\{\beta_{i} = (\tau_{i},\gamma_{i}, \mathbb{T}_{i}) \mid i \in \N\}$ is compatible, and $\mathfrak{B}_i = \{\beta_{i'} \mid i' \in D_i\}$ locally satisfies $\phi_i$, for all $i \in \N$.
\end{problem}

As each of the LTL formulas $\phi_i$, $i \in \{1,\ldots, N\}$ over~$\APs_i$ can be algorithmically translated into a \jana{deadlock-free} language equivalent BA~\cite{principles}, from now on, we pose the problem  with the task specification of each agent $i$  given as a  BA $\B_i = (Q_i, q_{\init, i}, \delta_i, \mathbf{\Sigma}_i=2^{\APs_i}, F_i)$ and the local task satisfaction condition formulated as $w_{\mathfrak B_i} \in L(\B_i)$. 

\begin{remark}
Collisions can be resolved either (i) through an LTL formula that forbids two agents to occupy the same cell of the environment or to exchange positions if they are in two neighboring cells, or (ii) through low-level controllers that implement the agents' transitions. This topic is however, beyond the scope of this paper.
\end{remark}

\label{sec:pf:cs}
A solution to Prob.~\ref{prob:main} can be obtained by imposing full synchronization and modifying the standard control plan synthesis procedure for TS from LTL specification ({see, e.g.,~\cite{marius-tac2008}):
(1) The  set of agents is first partitioned into dependency classes similarly as in \cite{meng-ijrr2015} by the iterative application of the rule that if $i' \in \D_i$, then $i'$ belongs to the same dependency class $I_\ell$ as $i$;
(2)~For each dependency class $I_\ell =\{{1_\ell},\ldots,{n_\ell} \}$ and each agent $i \in I_\ell$, we set $\gamma_{i}  = r_{i,1}r_{i,2} \ldots $, where $r_{i,j} = \sync_{i}$. This step yields compatible behaviors of all agents in $\N$ independently of their traces and transition time durations. 
(3) For each $I_\ell =\{ {1_\ell},\ldots,{n_\ell} \}$, a TS $\T_\ell$ with $S_\ell = S_{1_\ell}\times \ldots \times S_{n_\ell}$ is constructed that represents the stepwise-synchronized traces of the agents within the class. 
(4) A product $\P_\ell$ of $\T_\ell$ and $\B_{1_\ell}, \ldots, \B_{n_\ell}$ is built that captures only the agents' traces that are admissible by $\T_\ell$ and result into behaviors that locally satisfy $\phi_{1_\ell},\ldots,\phi_{n_\ell}$, respectively. The product $\P_\ell$ is analyzed using  graph algorithms to find its accepting run that projects onto the desired traces of $\T_i$, for all $i \in I_\ell$.
The outlined procedure is correct and complete.
Although a certain level of decentralization is achieved via dependency partition, the algorithm suffers from a exponential growth of the product automaton state space with the increasing number of dependent agents, making the approach infeasible in practice. 
Furthermore, the solution requires that the agents synchronize after every single action execution, which potentially  slows down the overall system performance. 

In this work, we aim to  reduce the  computational demands of the above solution and to prevent the unnecessarily frequent synchronization. 
We propose to decompose the infinite horizon planning problem into an infinite sequence of finite horizon planning problems that are solved iteratively, upon the execution of the system. 
We build the dependency classes dynamically at each iteration. These classes are then often smaller than the offline ones, which has a dramatic impact on the efficiency of the planning procedure. 
We show that the stepwise synchronization scheme can replaced with an event-triggered one. 
Finally, we prove, that under certain assumptions, the repetitive execution and recomputation of the plans leads to the  satisfaction of all specifications.

\section{Problem Solution}
\label{sec:solution}
This section provides the details of the proposed iterative solution to Prob.~\ref{prob:main}. 
\jana{In Sec.~\ref{sec:preliminary}, we set the preliminary synchronization sequences to be followed. In Sec.~\ref{sec:short} we present a finite horizon plan synthesis algorithm that consists of four steps: (1) partitioning the agents into classes based on their dependency; and then for each of the classes separately: (2) building an intersection specification automaton up to a predefined horizon; (3)~building a product  capturing system behaviors admissible by the all agents within the class and by the intersection specification automaton 
up to a predefined horizon and evaluating the states of the product to reflect their respective profit towards the satisfaction of the specifications; and (4) finding and projecting a path in the product that leads to the most profitable state onto finite trace fragments of the individual agents.} Sec.~\ref{sec:infinite} discusses the iterative execution and recomputation. 

Without loss of generality, let us assume that all agents in $\N$ form a single offline dependency class from Sec. \ref{sec:pf}. 

\subsection{Preliminary Synchronization Sequence}
\label{sec:preliminary}

We set $\gamma_{i}  = r_{i,1}r_{i,2} \ldots $, where $r_{i,j} = \sync_i$, i.e., the preliminary synchronization sequence ensures full synchronization of all the agents after every single action execution.
The behaviors of agents in $\N$ are thus compatible regardless of their traces and transition time durations. Namely, we have directly from Def.~\ref{def:behavior} and Def.~\ref{def:compatible}:

\begin{lemma} Let $\gamma_{i}  = r_{i,1}r_{i,2} \ldots $, where $r_{i,j} = \sync_i$, for all $i \in \N$. For  traces $\tau_{i} = s_{i,1}\alpha_{i,1}\ldots$, $\tau_{i'} = s_{i',1}\alpha_{i',1}\ldots$ of $\T_i,\T_{i'}$,  $i, i' \in \N$, it holds that $t_{i,\alpha_{j}} = t_{i',\alpha_{j}}$, for all $j \geq 1$.
\label{lemma:sync}
\end{lemma}

\subsection{Finite Horizon Planning}
\label{sec:short}
Besides the set of agents $\N = \{1, \ldots, N\}$ 
modeled as $\model_i = (\T_i, \Sync_i, \AP_i, L_i)$ 
and the specification BAs $\B_i = (Q_i, q_{\init, i}, \delta_i, \mathbf{\Sigma}_i=2^{\APs_i}, F)$, for all $i \in \N$, the inputs to the finite horizon planning algorithm are:
the current states of $\T_1,\ldots,\T_N$, denoted  $\currs_1, \ldots, \currs_N$, initially equal to $s_{\init,1},\ldots, s_{\init, N}$, respectively;
the current states of $\B_1,\ldots,\B_N$, denoted  $\currq_1, \ldots, \currq_N$, initially equal to $q_{\init,1},\ldots, q_{\init, N}$, respectively;
a linear ordering $\prec$ over $\N$, initially arbitrary;
 fixed horizons $\h, \H \in \Nat$, which, loosely speaking, set the depth of planning in each BA and TS, respectively.

\textbf{Dependency Partitioning.}
\label{sec:intersection}
We start with partitioning $\Spec = \{\B_1,\ldots, \B_N\}$ into the smallest possible subsets $\Phi_1,\ldots,\Phi_M$, such that none of the transitions of any $\B_i \in \Phi_\ell$ that appears within the horizon $\h$ from the current state~$\currq_i$ of $\B_i$ imposes restrictions on any agent $i'$, where $\B_{i'} \not \in \Phi_\ell$. This partition corresponds to the \emph{necessary and sufficient dependency} up to the horizon $\h$. 

\begin{definition}[Participating services] We call a set of services $\AP_{i'}$, $i' \in \D_i$  \emph{participating} in $q\in Q_i$ if 
$i'=i$, or 
there exist  $q'\in Q_i$,  $\sigma \in \Alpha_i$, and $\varsigma \subseteq \AP_{i'}$ such that $(q,\sigma, q') \in \delta_i$, and $(q, (\sigma \setminus \AP_{i'}) \cup \varsigma, q') \not \in \delta_i$.
\label{def:participating}

Given $q \in Q_i$, the \emph{alphabet $\Alpha_i^{\h}$ of $\B_i$ up to the horizon $\h$} is $\Alpha_i^{\h}(q) = 2^{\APs_i^h(q)}$, where 
$$\APs_i^h(q) = \bigcup_{\substack{q' \in \hat \delta_i^j(q), 0\leq j \leq \h}} \{\AP_{i'} \mid \AP_{i'} \text{ is participating in } q'\},$$
where $\hat \delta_i^j(q)$ denotes the set of states reachable from $q$ in $j$ steps (see Sec.~\ref{sec:prelims}).
\end{definition}

\begin{example}
An example of a BA is illustrated in Fig.~\ref{fig:BA_example}. Consider that this is a specification automaton $\B_1$, such that $\D_1 = \{1,2,3\}$, $\AP_1 = \{a\}$, $\AP_2 = \{b\}$, and $\AP_3 = \{c\}$, i.e., that $\APs_1 = \{a,b,c\}$, and $\Alpha_i = 2^{\{a,b,c\}}$. $\AP_1$ is by condition (i) of Def.~\ref{def:participating} participating in all states. $\AP_2$ is participating in $q_1$, since the service $b$ is required on the transition from $q_1$ to $q_2$. Formally, there exists $q' = q_2$, $\sigma = \{a,b\}$, and  $\varsigma = \emptyset$, such that $(q_1,\{a,b\},q_2) \in \delta_1$, but $(q_1, (\{a,b\} \setminus \{b\}) \cup \emptyset, q_2) \not \in \delta_1$. On the other hand, $\AP_3$ is not participating in $q_1$ as the service $c$ is neither required nor forbidden on the transition from $q_1$ to $q_2$. $\AP_3$ is participating in $q_2$ since $c$ is forbidden on the transition from $q_2$ to $q_3$ and moreover required on the transition from $q_2$ to $q_4$. Neither $\AP_2$ nor $\AP_3$ are participating in $q_4$. Hence, the alphabet up to the horizon $h=1$ and $h=2$ is $\APs_1^1(q_1) = \{a,b\}$ and $\APs_1^2(q_1) = \{a,b,c\}$, respectively.

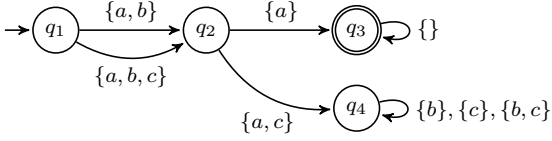
\begin{figure}[!h]
\scriptsize
\begin{center}
\usetikzlibrary{arrows,positioning,automata,shadows,fit,shapes}
\begin{tikzpicture}[->,>=stealth',shorten >=1pt,auto, node distance=2cm,semithick,initial text=]
  \tikzstyle{every state}=[draw]
	  \node[initial, initial where=left, state, minimum size=2ex] (q1) {$q_1$}; 
	  \node[state, minimum size=2ex] (q2) [right of = q1] {$q_2$}; 
	  \node[state, accepting, minimum size=2ex] (q3) [ right of = q2] {$q_3$}; 
	  \node[state, minimum size=2ex] (q4) [below =0.4cm of q3] {$q_4$}; 

	  \path (q1) edge node {$\{a,b\}$} (q2);
	  \path (q1) edge [bend right] node [below] {$ \{a,b,c\} $}  (q2);
	  \path (q2) edge node {$\{a\}$} (q3);
	  \path (q2) edge [bend right] node [below=0.1cm] {$\{a,c\}$} (q4);
	  \path (q3) edge [loop right] node [right] {$\{\}$} (q3);
	  \path (q4) edge [loop right] node [right] {$\begin{matrix} \{b\}, \{c\}, \{b,c\}\end{matrix}$} (q4);
\end{tikzpicture}
\end{center}
\normalsize
\caption{BA $\B_1$ with $\Alpha_1 = 2^{\{a,b,c\}}$. The states are illustrated as circles, the transitions as arrows labeled with the symbols from $\Alpha_1$. The arrow labeled with $ \{b\}, \{c\}, \{b,c\}$ represents 3 different transitions, under $\{b\}$, $\{c\}$, and $\{b,c\}$, respectively.}
\label{fig:BA_example}
\end{figure}
\end{example}

\begin{definition}[Dependency partition] 
\label{def:dep}
Given that $\currq_1, \ldots, \currq_N$ are the respective current states of BAs $\B_1,\ldots, \B_N$, the \emph{dependency partition} of $\Spec$ is induced by the equivalence $\sim^h$ defined on $\Spec$:
(i) $\B_i \sim^h \B_i$, and
(ii) if there exists $\B_{j}$, such that $\B_i \sim^h \B_{j}$, and $\AP_{i'} \subseteq \APs_{j}^h(\currq_{j})$ or $\AP_{j}  \subseteq \APs_{i'}^h(\currq_{i'})$, then also $\B_i \sim^h \B_{i'}$.
The dependency partition of $\Spec$ is then $\{ \Phi_1,\ldots, \Phi_M \}$, where $(\B_i \sim^h \B_{i'}) \iff (\B_i \in \Phi_\ell \iff \B_{i'} \in \Phi_\ell)$. 
The dependency partition of the set of all agents $\N$ is $\I = \{I_1,\ldots, I_M\}$, where a \emph{dependency class} $I_\ell$, is such that $ i \in I_\ell \iff \B_i \in \Phi_\ell$, for all $\ell \in \{1,\ldots, M\}$.
\end{definition}

The dependency partition is the function of the current states $\currq_1,\ldots,\currq_N$. Hence it is dynamically recomputed in each iteration of our solution. 
Although the local satisfaction of $\phi_i$ for  $i \in I_\ell$ depends on the traces of all the agents $j \in \D_i$, it is not influenced by the agents $j \in D_i \setminus I_\ell$ within the horizon~$h$. Thus, when we now concentrate on planning within the horizon $\h$, we can safely treat each {dependency class} $I_\ell = \{1_\ell, \ldots, n_\ell\}$ separately. In the worst case, the dynamic dependency partition equals to the offline one from Sec.~\ref{sec:pf:cs}. This is however, often not the case and even splitting a single offline dependency class into two online ones triggers an exponential improvement in terms of computational demands of further steps.
For an example, see Sec.~\ref{sec:simulations}.

\textbf{Intersection Automaton.}
For each dependency class $I_\ell$, we construct a finite automaton that represents the language intersection of the BAs in $\Phi_\ell = \{\B_{1_\ell},\ldots ,\B_{n_\ell}\}$ up to the pre-defined horizon $\h$. In this step, we rely on the fact that the preliminary synchronization sequence set in Sec.~\ref{sec:preliminary} guarantees step-by-step synchronization. We label the states of the intersection automaton with values that indicate the progress towards the satisfaction of the desired tasks.
Later on, these values are used to set temporary goals in the finite horizon plan synthesis. 

Without loss of generality, let the automata in $\Phi_\ell$ be ordered according to $\prec$, i.e. ${i_\ell} \prec {j_\ell}$, for all $1\leq i<j \leq n$. 

\begin{definition}[Intersection automaton]
\label{def:BA}
The intersection automaton of $\B_{1_\ell}, \ldots, \B_{n_\ell}$ up to the horizon $\h$ is $\A^h = (Q_\A, q_{\init, \A}, \Alpha_\A, \delta_\A, F_\A)$, where
$Q_\A \subset Q_{1_\ell} \times \ldots \times Q_{n_\ell} \times \Nat$ is a finite set of states, generated as described below;
$q_{\init, \A} = (\currq_{1_\ell}, \ldots, \currq_{n_\ell}, 1)$;
{$\Alpha_\A = \big \{ \bigcup_{i_\ell \in I_\ell} \sigma_{i_\ell} \mid \sigma_{i_\ell} \in (2^{\AP_{i_\ell}} \cup 2^{\Epsilon_{i_\ell}} ) \big \}$};
Let $Q_\A^0 = \{ q_{\init,\A} \}$. 
For all $1 \leq j \leq h$, we define $(q_{1_\ell}', \ldots, q_{n_\ell}', k') \in Q_\A^{j}$ and $\big((q_{1_\ell}, \ldots, q_{n_\ell}, k), \sigma, (q_{1_\ell}', \ldots, q_{n_\ell}', k')\big) \in \delta_\A^j$ iff (i) $(q_{1_\ell}, \ldots, q_{n_\ell}, k) \in Q_\A^{j-1}$, 
(ii) {for all $i_\ell \in I_\ell$, either$ (q_{i_\ell}, \sigma \cap \APs_{i_\ell} ,q_{i_\ell}') \in \delta_{i_\ell}$, or $q_{i_\ell} = q_{i_\ell}'$, and $\epsilon_{i_\ell} \in \sigma$, and (iii) $k' = k+1$ if $q_{{\kappa}_\ell}  \in F_{{\kappa}_\ell}$, where $\kappa = k \mod n_\ell$ and $k'=k$ otherwise.
Finally, $Q_\A = \bigcup_{0 \leq j \leq \h } Q_\A^j  \ \  \text{  and   } \  \ \delta_\A  = \bigcup_{1 \leq j \leq \h } \delta_\A^j; $}
$F_\A = \{ (q_{1_\ell},\ldots,q_{n_\ell}, k) \in Q_\A \setminus \{q_{\init, \A}\} \mid  q_{\kappa_\ell} \in F_{\kappa_\ell}, \text{ where } \kappa = k \mod n_\ell \}$.
\end{definition}

The intersection automaton is not interpreted over infinite words and hence, it is not a BA. However, it is an automaton and as such, it can be viewed as a graph (see Sec.~\ref{sec:prelims}). A~path in $\A^h$ from the initial state $(\currq_{1_\ell}, \ldots, \currq_{n_\ell}, 1)$ to a state $(q_{1_\ell},\ldots,q_{n_\ell}, k)$ corresponds to a path from $\currq_{i_\ell}$ to $q_{i_\ell}$ in each $\B_{i_\ell}$ and vice versa. Formally, these two lemmas follow from the construction:

\begin{lemma}
Consider a path $ q_{1} \sigma_1  q_2 \sigma_2 \ldots \sigma_{m-1}  q_m$ in $\A^h$, where $ q_j$ denotes the tuple $ ( q_{1_\ell,j},\ldots,  q_{n_\ell,j}, k_j) \in Q_\A$, for all $1 \leq j \leq m$. 
Let  $\varpi_{i_\ell,j} =  \sigma_j\cap ({\AP_{i_\ell}}\cup {\Epsilon_{i_\ell}})$ denote the range restriction of $\sigma_j$ to the services of agent $i_\ell$, for all $1 \leq j \leq m-1 $ and let $\varpi_{i_\ell,\i_1}\ldots \varpi_{i_\ell,\i_\mu}$ be the subsequence of non-silent elements of $\varpi_{i_\ell,1}\ldots\varpi_{i_\ell,m-1}$.
Finally, let $\omega_{i_\ell,j}  = \sigma_{\iota_j}  \cap {\APs_{i_\ell}}$, for all $1 \leq j  \leq \mu$.
Then there exists a \emph{corresponding run} $\bar \rho_{i_\ell} = \bar q_{i_\ell,1}\ldots \bar q_{i_\ell,{\mu}+1}\ldots$ of $\B_{i_\ell}$ over each word $\bar w_{i_\ell} = \omega_{i_\ell,1}\ldots \omega_{i_\ell,\mu}\ldots$, with the property that 
(1) $q_{i_\ell,1} = \bar q_{i_\ell,1} = \currq_{i_\ell}$,
(2)
$q_{i_\ell, \iota_j+1} = \bar q_{i_\ell, j+1}$, for all $1 \leq  j \leq \mu$,
(3) $q_{i_\ell, j+1} = q_{i_\ell, j}$, for all $1 \leq j \leq  m-1$, such that $j \neq \i_{j'}$, for any $1 \leq j' \leq  \mu$.
\label{lemma:correspondence1}
\end{lemma}

\begin{lemma}
Consider a run  $\rho_{i_\ell} = q_{i_\ell,1}q_{i_\ell,2}\ldots$ of $\B_{i_\ell}$ over a word $w_{i_\ell} = \omega_{i_\ell,1} \omega_{i_\ell,2}\ldots \in (2^{\APs_{i_\ell}})^\omega$, where $ q_{i_\ell,1} = \currq_{i_\ell}$.
Let $\sigma_1\sigma_2\ldots$ be a word over $\Alpha_\A$ with the property that there exists its subsequence $\sigma_{\i_1}\sigma_{\i_2}\ldots$, such that $\sigma_{\i_\iota} \cap {\APs_{i_\ell}} = \omega_{i_\ell,j} $, for all $j \geq 1$, while $\sigma_j \cap ({\AP_{i_\ell}}\cup {\Epsilon_{i_\ell}}) = \Epsilon_{i_\ell}$, for all $j \geq 1$, $j \neq \i_{j'}$, for any $j' \geq 1$. Then there exists a path $\bar q_{1} \sigma_1 \bar q_2 \sigma_2 \ldots \sigma_{h-1} \bar q_h$ in $\A^h$, where $\bar q_j$ denotes the tuple $(\bar q_{1_\ell,j},\ldots, \bar q_{n_\ell,j}, k_j) \in Q_\A$, such that: 
(1) $\bar q_{i_\ell,1}  = q_{i_\ell,1} = \currq_{i_\ell}$,
(2) $\bar q_{i_\ell,\iota_j+1} = q_{i_\ell,j+1}$, for all $1 \leq \iota_j \leq h$, and 
(3) $\bar q_{i_\ell,j+1} = \bar q_{i_\ell,j}$, for all $1 \leq j, \iota_{j'} < h$, where $j \neq \iota_{j'}$.
\label{lemma:correspondence2}
\end{lemma}

Through $k$, we remember the progress towards the satisfaction of the individual specifications ordered according to $\prec$; for $k \leq n_\ell$, an accepting state is guaranteed to be present on a the projected finite path of each $\B_{i_\ell}$, $1 \leq i \leq k$. For $k > n_\ell$, an accepting state is surely present on each projected run of each $\B_{i_\ell}$,  at least $\big \lceil  \frac{k}{n_\ell} \big \rceil$-times for all $1 \leq i \leq \kappa$, and at least $\big \lfloor \frac{k}{n_\ell} \big \rfloor$-times for all  $\kappa < i \leq n_\ell$, where $\kappa = k \mod n_\ell$.
To be able to identify ``profitable'' actions/transitions of the agents w.r.t.~$\Phi_\ell$, and thus also w.r.t. $\mathbf{\Phi}$, we assume that at least a state which represents a step towards the satisfaction of the highest-order specification $\B_{1_\ell}$ is present in $\A^h$. In Sec.~\ref{sec:product} this fact will allow us to set 
short-term goals in TSs $\T_{1_\ell}, \ldots, \T_{n_\ell}$.
We discuss its relaxation in Sec.~\ref{sec:discussion}.

\begin{assumption} 
\label{assump:BA}
Assume that $F_\A$ is not empty. 
\label{assump:1}
\end{assumption}


\begin{definition}[Automaton progressive function]
\label{def:VB} The progressive function $V_{\A}: Q_\A \to \Nat_0 \times \mathbb{Z}_0^-$ is for a state $q=(q_{1_\ell}, \ldots, q_{n_\ell}, k)$ defined as:
$V_{\A}(q) =  \big(k,-\min_{q_f \in F_\A} \dist(q,q_f)\big).$
\end{definition}

The increasing value of $V_{\A}$ indicates a progress towards the satisfaction of the individual local specifications in $\Phi_\ell$, ordered according to $\prec$. No progress can be achieved from a state $q$, such that $V_{\A}(q) = (k,-\infty)$ within the horizon $\h$, and hence, we remove such states 
from $\A^h$. From Assump.~\ref{assump:BA}, $V_{\A}(q_{\init,\A}) = (1, d)$, where $d\neq -\infty$.

\textbf{Product System.}
The intersection automaton and its progressive function allow us to assess which service sets should be provided in order to maximize the progress towards the satisfaction of the specifications. The remaining step is to plan the transitions of the individual agents to reach states in which these services are available. We do so through the definition of a product system $P^H$. Besides the behaviors permitted by the task specifications,  the product system captures the allowed behaviors (finite trace fragments) of the agents from $I_\ell$ up to the horizon $\H$. Similarly as the states of $A^h$, the states of $P^H$ are evaluated to indicate their progress towards the specifications satisfaction.

\label{sec:product}
\begin{definition}[Product system]
\label{def:PA}
The product system up to the horizon $\H$ of the agent TSs $\T_{i_\ell}, i_\ell \in I_\ell$, and the intersection automaton $\A^h$ from Def.~\ref{def:BA} is an automaton $\P^\H= (Q_\P, q_{\init,\P},  \Alpha_\P, \delta_\P)$, where $Q_\P \subset S_{1_\ell} \times \ldots \times S_{n_\ell} \times Q_\A$ is a finite set of states, generated as described below;
$q_{\init,\P} = (\currs_{1_\ell},\ldots,\currs_{n_\ell},q_{\init,\A})$;
$\Alpha_\P = A_{1_\ell} \times \ldots \times A_{n_\ell} \times \Alpha_\A;$
Let $Q^0_\P = \{ q_{\init,\P} \}$. 
For all $1 \leq j \leq H$, $q = (s_{1_\ell},\ldots,s_{n_\ell},q_\A)$, $\sigma =  (\alpha_{1_\ell},\ldots,\alpha_{n_\ell},\sigma_\A)$, $q' = 
(s_{1_\ell}',\ldots, s_{n_\ell}',q'_\A)$, we define that
$q' \in Q_\P^{j}$ and
$(q,\sigma,q') \in \delta_\P^j$
iff 
(i) $q \in Q_\P^{j-1}$, 
(ii){for all $i_\ell \in I_\ell$, it holds that $T(s_{i_\ell},\alpha_{i_\ell}) =  s_{i_\ell}'$, and $\sigma_\A \cap (\AP_{i_\ell} \cup \Epsilon_{i_\ell} ) = L_{i_\ell}(\alpha_{i_\ell})$, and}
(iii)$(q_\A,\sigma_\A,q_\A') \in \delta_\A$.
Finally, $Q_\P = \bigcup_{0 \leq j \leq H} Q^j_\P \ \ \text{ and } \ \ \delta_\P  = \bigcup_{1 \leq j \leq H} \delta^j_\P. $
\label{def:product}
\end{definition}

The set of accepting states $F_\P$ is not significant for the further computations, hence we omit it from $\P^\H$. Analogously as $\A^h$, $\P^\H$ can be viewed as a graph (see Sec.~\ref{sec:prelims}). A path in $\P^H$ can be projected onto a finite trace prefix of each $\T_{i_\ell}$, a finite run prefix of $\A^h$ and further through Lemma~\ref{lemma:correspondence1} onto a finite run prefix of each BA $\B_{i_\ell}$, too. 

\begin{definition}[Projection]
Consider a  path ${\varrho} = q_1{\sigma_1}q_2 \ldots q_{m-1}{\sigma_{m-1}}q_m$ in $\P^\H$, where $q_1 = q_{\init,\P}$. $\varrho$ can be projected onto a finite trace  prefix $\hat \tau_{i_\ell}(\varrho)$ of each $\T_{i_\ell}$, $i_\ell \in I_\ell$ in the expected way: for all $1 \leq j \leq m$, the $j$-th state and action of $\hat \tau_{{i_\ell}}(\varrho)$ is $s_{i_\ell,j}$ and $\alpha_{i_\ell,j}$ if the $j$-th state and transition label of $\varrho$ is $q_j=(s_{1_\ell,j}, \ldots, s_{n_\ell,j},q_{\A,j})$, and $\sigma_{j}=(\alpha_{1_\ell,j}, \ldots, \alpha_{n_\ell,j},\sigma_{\A,j})$, respectively.
Furthermore, for all $1 \leq j \leq m$, the $j$-th state of the projected run prefix $\hat \rho_{\A}(\varrho) $ of $\A^h$ is $q_{\A,j}$ if the $j$-th state of $\varrho$ is $q_j = (s_{1_\ell},\ldots,s_{n_\ell},q_{\A,j})$; and assuming that $q_{\A,j}= (q_{1_\ell,j},\ldots, q_{n_\ell,j}, k)$, the state of the corresponding state sequence $\hat \rho_{{i_\ell}}(\varrho)$ in $\B_{i_\ell}$ is $q_{i_\ell,j}$.
\label{def:proj}
\end{definition}

Although $\hat \rho_{{i_\ell}}(\varrho)$ is a projection of $q_{\A,j}$, i.e., a sequence of states of $\B_{i_\ell}$, it might not be a run of $\B_{i_\ell}$ as such. However, thanks to Lemma~\ref{lemma:correspondence1}}, $\hat \rho_{{i_\ell}}(\varrho)$ maps to a unique corresponding run $\bar{\rho}_{i_\ell}(\varrho)$ of $q_{\B_{i_\ell}}$. Note that in what follows, we distinguish between  $\hat \rho_{{i_\ell}}(\varrho)$ and $\bar{\rho}_{i_\ell}(\varrho)$.

\begin{definition}[Progressive function and state] The progressive function $V_{\P}: Q_\P \to \Nat_0 \times \mathbb{Z}_0^-$ is inherited from the intersection automaton $\A^h$ (Def. \ref{def:VB}), \ie for all $q = (s_{1_\ell},\ldots, s_{n_\ell},q_\A) \in Q_\P$,
$V_\P\big((s_{1_\ell},\ldots, s_{n_\ell},q_\A)\big)= V_{\A}(q_\A).$
A  state $q\in Q_\P$ is a \emph{progressive state} if $V_\P(q) > V_\P(q_{\init, \P})$. A \emph{maximally progressive state} is a progressive state $q$, with the property that for all $q' \in Q_\P$, it holds $V_\P(q) \geq V_\P(q')$. 
\end{definition}

\textbf{Finite Horizon Plan Synthesis.}
\label{sec:plan}
To find a suitable finite horizon plan, we impose the following assumption and discuss its relaxation in Sec.~\ref{sec:discussion}. 
It states that within $H$, at least one service set can be provided that makes at least one step towards the local satisfaction of the highest-priority formula $\phi_{1_\ell}$. 

\begin{assumption}
\label{assump:PA}
Assume that there is a progressive state $q_\mathit{p}$ reachable in $\P^H$ via a finite path $q_1\sigma_1 \ldots \sigma_{m-1}q_m$, where $q_1  = q_{\init,\P}$, $q_m = q_p$, and $L(\alpha_{1_\ell,j}) \neq \Epsilon_{1_\ell,j}$, for some $\sigma_j =  (\alpha_{1_\ell,j},\ldots,\alpha_{n_\ell,j},\sigma_{\A,j})$, $1 \leq j \leq m-1$.
\end{assumption}

We compute a suitable finite horizon plan as the shortest path $\varrho = q_1{\sigma_1}q_2 \ldots q_{m}{\sigma_{m}}q_{\mathit{m}}$ in $\P^H$ from $q_1=q_{\init,\P}$ to a maximally progressive state~$q_m = q_{\mathit{max}}$.
Such a path can be found in linear time with respect to the size of~$\P^\H$ (see~Sec.~\ref{sec:prelims}).
The projection of $\varrho$ onto the individual  TSs gives finite trace fragments $\hat \tau_{i_\ell}(\varrho) = \currs_{i_\ell} \alpha_{i_\ell,1} s_{i_\ell,2} \ldots s_{i_\ell, m-1} \alpha_{i_\ell, m-1} s_{i_\ell,m}$, to be followed by each agent $i_\ell \in I_\ell$. 
Alg.~\ref{alg:short} summarizes the proposed finite horizon planning for the set of all agents $\N$. Since this algorithm will be run iteratively, starting from its second execution on, it also takes as an input  the  fragments $\hat \tau_1, \ldots, \hat \tau_N, \hat \rho_1, \ldots, \hat \rho_N$ and the progressive function value of the maximally progressive state from the previous iteration. In comparison to the solution from Sec.~\ref{sec:pf:cs}, we treat separately each dynamic dependency class (lines~\ref{line:2}--\ref{line:9}). These classes are often smaller than the offline dependency classes, 
thereby increasing the level of decentralization of the planning procedure. 

{
\begin{algorithm}[!h]
\caption{Procedure $\mathsf{short\_horizon\_plan}$}
\label{alg:short}
\begin{algorithmic}[1]
\small
\INPUT a set of agents $\N = \{1,\ldots,N\}$, their models $\model_1,\ldots, \model_N$; BAs $\B_1,\ldots, \B_N$;  current states $\currs_1, \ldots, \currs_n$, $\currq_1,\ldots, \currq_n$; an ordering $\prec$ over $\N$; and planning horizons $\h, H \in \Nat$, previous fragments  $\hat \tau_1, \ldots, \hat \tau_N, \hat \rho_1, \ldots, \hat \rho_N$, previous maximal progressive function value $V_{\P,max}$
\OUTPUT 
finite trace fragments $\hat \tau_1,\ldots, \hat \tau_N$ of $\T_1,\ldots, \T_N$; finite state sequences $\hat \rho_1,\ldots,\hat \rho_N$ of $\B_1,\ldots, \B_\N$; and a maximal progressive function value $V_{\P,max}$
\STATE compute the partition $\I= \{I_1,\ldots,I_M\}$ (Def.~\ref{def:dep})
\FORALL{ $\ell \in \{1,\ldots,M\}$} \label{line:2}
\STATE construct $\A^h$ (Def.~\ref{def:BA}) and construct $\P^H$ (Def.~\ref{def:PA})
\STATE find a maximally progressive state $q_{max}$ in $\P^\H$ \label{line:5}
\IF {$V_\P(q_{max}) > V_{\P,max}$} \label{line:6}
\STATE find the shortest path $\varrho$ to $q_{max}$ and update $V_{\P,max}$
\FORALL {$i_\ell \in I_\ell$} \label{line:8a}
\STATE $\hat \tau_{i_\ell} := \hat \tau_{i_\ell} (\varrho)$ and  $\hat \rho_{i_\ell} := \hat \rho_{i_\ell}(\varrho)$ (Def.~\ref{def:proj})
\ENDFOR\label{line:8}
\ELSE 
\STATE remain $\hat \tau_1, \ldots, \hat \tau_N, \hat \rho_1, \ldots, \hat \rho_N$, $V_{\P,max}$ unchanged
\ENDIF
\ENDFOR\label{line:9}
\RETURN $
\hat \tau_1, \ldots, \hat \tau_N, \hat \rho_1, \ldots, \hat \rho_N$, $V_{\P,max}$
\end{algorithmic}
\end{algorithm}
}

\subsection{Infinite Horizon Replanning}
\label{sec:infinite}
To complete the solution, we discuss the infinite, iterative execution and recomputation of the finite horizon plans. We present two  recomputation strategies: a \emph{stepwise} one, and later on in Sec.~\ref{sec:event} an \emph{event-triggered} one, where we adjust the preliminary synchronization sequences to reduce the synchronization frequency.

\subsubsection{Stepwise Solution}

In each iteration of the stepwise solution, the finite horizon plans are executed as summarized in Alg.~\ref{alg:main1}. Each agent $i \in \N$ first sets its preliminary synchronization sequence $\gamma_i = r_{i,1}r_{i,2}\ldots$  and synchronizes with the other agents, i.e., sends $r_{i,1} = \sync_i$, and waits for the reception of $\sync_{i'}$ from all $i' \in \N$. Second, it computes the finite horizon plans by Alg.~\ref{alg:short}.
Third, it
executes the first action $\alpha_{i,1}$ of the planned trace fragment $\hat \tau_{i} = \currs_{i} \alpha_{i,1} s_{i,2} \ldots s_{i, m-1} \alpha_{i, m-1} s_{i,m}$ along with the first silent or non-silent service set $L(\alpha_{i,1})$. It transitions to state $s_{i,2}$ and the current state $\currs_{i}$ is updated.
The previous synchronization through $r_{i,1}$ has ensured that each agent  $i' \in I_\ell$ executes its respective action $\alpha_{i',1}$ simultaneously. 
At the same time, the current state of the BAs $\B_i$, is updated to the second state $q_{i,2}$ of the run fragment $\hat \rho_i$. Note, that if $L(\alpha_{i,1}) = \epsilon_{i}$, then $q_{i,2} = q_{i,1}$ by Def.~\ref{def:BA} and Def.~\ref{def:product}. Furthermore, if $q_{i,2} \in F_i$, then we update the ordering $\prec$ so that $i$ becomes of the lowest priority (the highest order), 
while maintaining the mutual ordering of all the other agents. 
This change reflects that a step towards the local satisfaction of $\B_i$ has been made and in the next iteration, progressing towards another agent's specification will be prioritized. Finally, all the agents 
start another iteration of the algorithm simultaneously 
as prescribed by the synchronization sequence. For the simplicity of the presentation, we assumed that the computation of  $\mathsf{short\_horizon\_plan}$ does not take any time. In practice, we would cope with different computation times via synchronization both before and after running the procedure $\mathsf{short\_horizon\_plan}$.

\begin{algorithm}[!h]
\caption{Stepwise solution to Prob.~\ref{prob:main}}
\label{alg:main1}
\begin{algorithmic}[1]
\small
\INPUT a set of agents $\N = \{1,\ldots,N\}$, their models $\model_1,\ldots, \model_N$; BAs $\B_1,\ldots, \B_N$; and  horizons $\h, H \in \Nat$
\OUTPUT traces $\tau_1,\ldots, \tau_N$ and synchronization sequences $\gamma_1,\ldots, \gamma_N$ that are a solution to Problem~\ref{prob:main}, and sequences $\rho_1,\ldots,\rho_n$ of states of BAs $\B_1,\ldots, \B_N$.
\FORALL {$i\in \N$} 
\STATE initialize $\prec\,:=(1,\ldots,N)$; $\currs_i:=s_{\init,i}$ $\currq_i := q_{\init,i}$
\STATE initialize $\hat{\tau}_i := $ empty; $\hat{\rho}_i := $ empty; $V_{\P,max} := (0,0)$
\STATE initialize synchronization sequence $\gamma_i$ (Sec.~\ref{sec:preliminary})
\STATE  send $r_{i,1} := \sync_i$ and wait for $\sync_{i'}$ from all $i' \in \N$
\STATE $j_i:= 2$
\ENDFOR
\WHILE {true}
\STATE $
\hat \tau_1, \ldots, \hat \tau_N, \hat \rho_1, \ldots, \hat \rho_N, V_{\P,max}$ : =  $\mathsf{short\_horizon\_plan}$ \label{line:2:8}
\FORALL {$i\in \N$}
\STATE  execute $\alpha_{i,1}$, provide $L(\alpha_{i,1})$; $\currs_i :=  s_{i,2}$; $\currq_i := q_{i,2}$
\IF {$\currq_i \in F_i$} \label{line:12}
\STATE reorder $\prec$, s.t. $i' \prec i$, for all $i' \in \{1,\ldots,N\} \setminus \{i\}$
\STATE $V_{\P,max} := (0,0)$
\ENDIF \label{line:14}
\STATE send $r_{i,j_i} := \sync_i$, wait for $\sync_{i'}$ from all $i' \in \N$ 
\STATE $j_i:= j_i + 1$
\ENDFOR
\ENDWHILE
\RETURN $\tau_1,\ldots,\tau_N$, $\gamma_1,\ldots,\gamma_n$, $\rho_1,\ldots,\rho_n$
\end{algorithmic}
\end{algorithm}

\jana{The motivation for replanning after every iteration on line 11 of Alg.~\ref{alg:main1} is to gradually shift the planning horizon in order to keep the planning procedure sufficiently informed about the future possibilities. Less frequent replanning may be more efficient in terms of computational demands, at the cost of inefficiency of the resulting plans. Although the TSs and the BAs are finite and hence there are a finite number of different finite horizon problems to solve, the considered LTL formulas are interpreted over infinite time and the number of iterations cannot be generally upper-bounded due to the arbitrary transition durations. An alternative to the infinite number of executions would be to remember how the finite horizon problem has been resolved for every single  combination of the TSs and BAs states; however, this is generally not feasible due to the fact that the number of these combinations grows quickly with the number of agents, the size of the environment and the complexity of the tasks.}

\subsubsection{Event-Triggered Solution}
\label{sec:event}
In the stepwise solution, the synchronization takes place after every single transition of every agent, which might be more frequently than necessarily needed. For example, if the first $m$ actions in the planned trace fragment $\hat \tau_i$ of an agent $i$ are all associated with silent services, then 
this agent 
does not need to synchronize with the others nor to recompute its planned trace fragment. In what follows, we adapt the preliminary synchronization sequence so that the synchronization and recomputation are triggered  by the need of collaboration.

Assume that for agent $i$ the procedure $\mathsf{short\_horizon\_plan}$ executed on line~\ref{line:2:8} of Alg.~\ref{alg:main1} has returned a trace and a run fragment denoted by $\hat \tau_i = s_{i,1}\alpha_{i,1}s_{i,2}\alpha_{i,2} \ldots s_{i,m}$ and $\hat \rho_{i} = q_{i,1}q_{i,2}\ldots q_{i,m}$, for some $m \geq 2$. The main idea is to postpone the synchronization with the others from after the execution of $\alpha_{i,1}$ till the time $t_{s_{i,j}}$ right before the execution of the action $\alpha_{i,j}, j >1$ with one of the following properties:
$L(\alpha_{i,j})$ is non-silent, or
$q_{i,j}$ is accepting, or
$q_{i,j} = q_{i,m}$, or
there exists an agent $i'$, such that $\sync_{i'}$ has been received during $(t_{\alpha_{i,j-1}},t_{s_{i,j}}]$.
If one of the conditions is met, the agent sends $\sync_i$, and waits for receiving $\sync_{i'}$, from all ${i'} \in \N$. The finite horizon plans are recomputed and the event-triggered recomputation procedure repeats. On the other hand, if none of the conditions is met, the agent $i$ substitutes the synchronization action $\sync_i$ planned within the preliminary synchronization sequence $\gamma_i$ with $\nosync_i$. \jana{Note that thanks to the enforced compatibility of the behaviors, a deadlock is prevented.} The solution is summarized in Alg.~\ref{alg:main2}.

\begin{algorithm}[ht]
\caption{Event-triggered solution to Prob.~\ref{prob:main}}
\label{alg:main2}
\begin{algorithmic}[1]
\small
\INPUT a set of agents $\N = \{1,\ldots,N\}$, their models $\model_1,\ldots, \model_N$; BAs $\B_1,\ldots, \B_N$; and horizons $\h, H \in \Nat$
\OUTPUT traces $\tau_1,\ldots, \tau_N$ and synchronization sequences $\gamma_1,\ldots, \gamma_N$ that are a solution to Problem~\ref{prob:main}, and sequences $\rho_1,\ldots,\rho_n$ of states of BAs $\B_1,\ldots, \B_N$.
\FORALL {$i\in \N$} 
\STATE initialize $\prec\,:=(1,\ldots,N)$; $\currs_i:=s_{\init,i}$ $\currq_i := q_{\init,i}$
\STATE initialize $\hat{\tau}_i := $ empty; $\hat{\rho}_i := $ empty; $V_{\P,max} := (0,0)$
\STATE initialize synchronization sequence $\gamma_i$ (Sec.~\ref{sec:preliminary})
\STATE  send $r_{i,1} := \sync_i$ and wait for $\sync_{i'}$ from all $i' \in \N$
\STATE $j_i:= 2$
\ENDFOR
\WHILE {true}
\STATE $
\hat \tau_1, \ldots, \hat \tau_N, \hat \rho_1, \ldots, \hat \rho_N$ : =  $\mathsf{short\_horizon\_plan}$ \label{line:alg3:9}
\FORALL {$i\in \N$}
\STATE  execute $\alpha_{i,1}$ and provide service set $L(\alpha_{i,1})$
\STATE $\currs_i :=  s_{i,2}$; $\currq_i := q_{i,2}$; $k_i := 2$
\WHILE {$L(\alpha_{i,j_i}) = \epsilon_i$ and $\currq_i \not \in F_i$ and $\currq_i$ is not the last element of $\hat \rho_i$ and $\sync_{i'}$ was not received from any $i' \in \N \setminus \{i\}$ during the last iteration} \label{line:2:14}
 \STATE  send $r_{i,j_i} := \nosync_i$; $j_i:= j_i+1$
\STATE  execute $\alpha_{i,k_i}$ and provide service set $L(\alpha_{i,k_i})$
\STATE $\currs_i :=  s_{i,k_i+1}$; $\currq_i := q_{i,k_i+1}$; $k_i := k_i +1$
\ENDWHILE 
\IF {$\currq_i \in F_i$} \label{line:2:20}
\STATE reorder $\prec$, s.t. $j \prec i$, for all $j \in \{1,\ldots,N\} \setminus \{i\}$
\STATE $V_{\P,max} := (0,0)$
\ENDIF \label{line:2:22}
\STATE send $r_{i,j_i} := \sync_i$, wait for $\sync_{i'}$ from all $i' \in \N$
\STATE $j_i:= j_i + 1$
\ENDFOR
\ENDWHILE
\RETURN $\tau_1,\ldots,\tau_N$, $\gamma_1,\ldots,\gamma_n$, $\rho_1,\ldots,\rho_n$
\end{algorithmic}
\end{algorithm}

\begin{remark}
In Remark~\ref{remark:sync1}, we introduced parametrized synchronization requests. Assume that $\I = \{I_1,\ldots, I_M\}$ is an offline dependency partition from Sec.~\ref{sec:pf:cs}. Then for each agent $i \in I_\ell$, we can  replace $\sync_i$ with $\sync_i(I_\ell)$, and every $\nosync_i$ with $\sync_i(\{i\})$. However, an analogous step cannot be applied in the case of dynamic classes.
\label{remark:sync2}
\end{remark}

\section{\jana{Solution Analysis and Discussion}}
\label{sec:discussion}
\begin{lemma}
Let $\tau_1,\ldots,\tau_N$, $\gamma_1,\ldots,\gamma_n$, $\rho_1,\ldots,\rho_n$ be the traces, the synchronization sequences and the corresponding state sequences returned by Alg.~\ref{alg:main1}. Then $\rho_i$ contains infinitely many states $f_i \in F_i$, for all $i \in \N$.
\label{lemma:accepting}\\
\textbf{\emph{Proof.}} 
Denote $\tau_i = s_{i,1}s_{i,2}\ldots$, $\rho_i = q_{i,1}q_{i,2}\ldots$, and $\mathbb{T}(\tau_i) = t_{s_{i,1}}t_{s_{i,2}} \ldots$, for all $i \in \N$. Consider a time instance $t_{s_{i,j}}$, where $j\geq 1$ and assume that $i$ is the most prioritized agent at  $t_{s_{i,j}}$, i.e., that $i \prec i'$, for all $i' \in \N$. Let $\varrho$ be the  path  to a maximally progressive state $q_\mathit{max}$ of $\P^H$ at time $t_{s_{i,j}}$ computed on line~\ref{line:5} of Alg.~\ref{alg:short} and let $\hat{\tau_i}$ and $\hat{\rho_i}$ be the corresponding finite trace prefix of $\T_i$, and the state sequence of $\B_i$ computed on lines~\ref{line:8a}--\ref{line:8}. 
Assume for a moment that the dependency partition $\I$ remains the same at time $t_{s_{i,j+1}}$.
Regardless of the steps taken in Alg.~\ref{alg:main1}, the state $q_\mathit{max}$ is also present in $\P^H$ at time $t_{s_{i,{j+1}}}$, hence if condition on line~\ref{line:6} is not satisfied, this state remains the progressive goal. Now, let $\varrho'$ be the  path  to a maximally progressive state $q_\mathit{max}'$ of $\P^H$ at time $t_{s_{i,j}+1}$ and let $\hat{\tau_i}'$ and $\hat{\rho_i'}$ be the corresponding finite trace prefix of $\T_i$, and the state sequence of $\B_i$. If $q_\mathit{max}' \neq q_\mathit{max}$, then $V_\P(q_\mathit{max}') >  V_\P(q_\mathit{max})$ and $q_\mathit{max}'$ is, loosely speaking,  closer to reaching an accepting state of $\B_i$. Thanks to  Assump.~\ref{assump:BA} and Assump.~\ref{assump:PA} and  the finite number of states of $\P^\H$, by repetitive reasoning we get that there exists time $t_{s_{i,m}}$, when a state $q^*$ of $\P_H$ is reached, such that $V_\P(q^*)\geq V_\P(q_\mathit{max})$. Inductively, a state $q_f$ of $P^\H$ that projects onto an accepting state of $\B_i$ will eventually be reached. 
Similar holds even if the dependency partition $\I$ has changed at time $t_{s_{i,j+1}}$ and $i$ is present in a dependency class $I_{\ell'} \neq I_\ell$. The state $q_\mathit{max} = (s_{1_\ell,m},\ldots, s_{n_\ell, m}, (q_{1_\ell,m},\ldots,q_{n_\ell,m}, k_m))$ can be mapped onto a state of the new $\P^H$ at time $t_{s_{i,j+1}}$, i.e., onto $q_\mathit{max, \ell'} = (s_{1_{\ell'},m'},\ldots, s_{n_{\ell'}, m'}, (q_{1_{\ell'},m'},\ldots,q_{n_{\ell'},m'}, k_{m'}))$, where $s_{i',m} = s_{i',m'}$, and $q_{i',m} = q_{i', m'}$, for all $i' \in I_\ell \cap I_{\ell'}$. The remainder of the proof is analogous to the above.
Finally, it is ensured that at least one non-silent service is provided by $\T_i$ on the executed path to $q_f$. Lines~\ref{line:12}--\ref{line:14} of Alg.~\ref{alg:main1} ensure that each $i \in \N$ will repeatedly become the most prioritized one. Altogether, $\phi_i$ contains infinitely many states $f_i \in F_i$, for all $i \in \N$.
\end{lemma}

In summary, Lemma~\ref{lemma:sync} gives us the existence of compatible behaviors regardless of the traces of the agent TSs and the transition time durations. Lemma~\ref{lemma:accepting} together with Lemmas~\ref{lemma:correspondence1}, and~\ref{lemma:correspondence2}, and Def.~\ref{def:PA} yield that $\tau_1,\ldots,\tau_N$ produce words that are accepted by each $\B_i$.

\begin{theorem}
The traces $\tau_1,\ldots, \tau_N$ together with the synchronization sequences $\gamma_1,\ldots, \gamma_N$ returned by Alg.~\ref{alg:main1} provide a solution to Prob.~\ref{prob:main}. 
\end{theorem}

To prove the correctness of the event-triggered solution, we have to prove that the computed traces and synchronization sequences (i)~yield compatible behaviors and (ii)~locally satisfy the LTL formulas:
\begin{lemma}
The traces $\tau_1,\ldots, \tau_N$,  where $\tau_i = s_{i,1}\alpha_{i,1}\ldots$, for all $i \in \N$,  together with the synchronization sequences $\gamma_1,\ldots, \gamma_N$ returned by Alg.~\ref{alg:main2} yield compatible behaviors regardless of the values of the transition time durations $\Delta_{\alpha_{i,1}}, \Delta_{\alpha_{i,2}}\ldots$.\\
\textbf{\emph{Proof.}}
{Follows immediately from the condition of the while loop on line~\ref{line:2:14} of Alg.~\ref{alg:main2}}.
\end{lemma}

\begin{lemma}
Let $\tau_1,\ldots,\tau_N$, $\gamma_1,\ldots,\gamma_N$, $\rho_1,\ldots,\rho_N$ be the traces, the synchronization sequences and the corresponding state sequences returned by Alg.~\ref{alg:main2}. Then $\rho_i$ contains infinitely many states $f_i \in F_i$, for all $i \in \N$.
\textbf{\emph{Proof.}}
Denote $\tau_i = s_{i,1}\alpha_{i,1}\ldots$, $\rho_i = q_{i,1}q_{i,2}\ldots$, and $\mathbb T_i= t_{s_{i,1}}t_{\alpha_{i,1}} t_{s_{i,2}}t_{\alpha_{i,2}} \ldots$, for all $i \in \N$.
To prove the lemma, we prove that at each time $t$, it holds that an accepting state $f_i \in F_i$ will be eventually reached for the most prioritized agent $i$ at time $t$.  First, consider $t=0$. Without loss of generality, assume that $i \prec i'$, for all $i' \in \N$. Denote $I_\ell \in \I$ the dependency class $i$ belongs to. 
Let $\hat{\tau}_1, \ldots, \hat{\tau}_N$ and $\hat{\rho}_1, \ldots ,\hat{\rho}_N$ be the finite trace prefixes of $\T_1, \ldots, \T_N$, and the state sequences of $\B_1,\ldots, \B_N$ computed on line~\ref{line:alg3:9} of Alg.~\ref{alg:main2}, respectively. 
Specifically for $I_\ell$, these were obtained from the projections of the shortest path $\varrho_\ell$ that leads to a maximally progressive state $q_\mathit{max,\ell} = (s_{1_\ell,m}\ldots s_{n_\ell,m},(q_{1_\ell,m},\ldots,q_{n_\ell,m} ,k_m))$ of the product system  $\P^H_\ell$ computed on line~\ref{line:5} of Alg.~\ref{alg:short}.
Let the execution proceed as described in Alg.~\ref{alg:main2} and let $\mathfrak{t}_{s}$ denote the first time instance after $t = 0$ with one of the properties triggering the synchronization, i.e., with one of the conditions of line 13 of Alg.~\ref{alg:main2} being false. Note that $\mathfrak{t}_{s}$ is finite and after the requested synchronization is performed, the time $\mathfrak{t}_{{\alpha}}$ equals to some $t_{s_{i',j}}$ in $\mathbb T_{i'}$, for all $i'\in \N$.
Let $\currs_{i'}$ and $\currq_{i'}$ denote the respective states of $\T_{i'}$ and $\B_{i'}$ at time $\mathfrak t_\alpha$, and let $\vec{\tau}_{i'}$ and $\vec{\rho}_{i'}$ denote the respective suffixes of $\hat{\tau}_{i'}$ and $\hat{\rho}_{i'}$ starting at $\currs_{i'}$ and $\currq_{i'}$, for all $i' \in \N$.  
Denote by ${I}_{\ell'}$ the dependency class $i$ belongs to at time $\mathfrak t_\alpha$ and by $\P^H_{\ell'}$ the corresponding product system. 
We now show the existence of a mapping between the suffixes $\vec{\tau}_{i'}$ and $\vec{\rho}_{i'}$ of agents $i' \in {I}_{\ell'} \cap I_\ell$  onto a single finite path ${\varrho}_{\ell'}$ in $\P^H_{\ell'}$, whose length is strictly smaller than the length of $\varrho_\ell$ in $\P^H_\ell$. 
Trivially, if $s_{i',m} = \currs_{i'}$, for all $i' \in {I}_{\ell'} \cap I_\ell$, the path $\varrho_{\ell'}$ is empty. Suppose that  $s_{i',m} \neq \currs_{i'}$, for some $i' \in {I}_{\ell'} \cap I_\ell$.
We propose $\varrho_{\ell'}$ to be the path whose projections onto the states of $\T_{i'}$ and $\B_{i'}$ are the respective sequences $\currs_{i'} \ldots \currs_{i'}\vec{\tau}_{i'}$ and $\currq_{i'}\ldots \currq_{i'}\vec{\rho}_{i'}$, such that the length of these two sequences, and hence also the length of $\varrho_{\ell'}$, equals to the length of the  suffix $\vec{\tau}_{i''}$ that is the longest one among the agents agents $i'' \in {I}_{\ell'} \cap I_\ell$. This path is  strictly shorter than $\varrho_\ell$. At the same time, it exists in $\P^H_{\ell'}$ due to the assumption that $s \xrightarrow{\alpha} s$, for all $s \in S_{i'}$, $i' \in \N$, and some $\alpha \in A_{i'}$. The existence of $\varrho_{\ell'}$ ensures that a progress towards some $f_i \in F_i$. 
Thanks to  Assump.~\ref{assump:BA} and Assump.~\ref{assump:PA} and the finite  number of states of the the product system, by repetitive reasoning we get that during the execution of Alg.~\ref{alg:main2}, since a certain moment on, the last state of $\varrho_\ell'$ will be the maximally progressive state $q_\mathit{max,\ell'}$ in $\P^H_{\ell'}$, until this state is reached. Inductively, a state $q_f$ that projects onto an accepting state of $\B_i$ will eventually be reached. At the same time, at least one non-silent service is provided by $\T_i$ on the path to $q_f$. Furthermore, lines~\ref{line:2:20}--\ref{line:2:22} of Alg.~\ref{alg:main2} ensure, that each $i \in \N$ will repeatedly become the most prioritized one. Altogether, we $\rho_i$ contains infinitely many states $f_i \in F_i$, for all $i \in \N$.
\end{lemma}

\begin{theorem}
The traces $\tau_1,\ldots, \tau_N$ together with the synchronization sequences $\gamma_1,\ldots, \gamma_N$ returned by Alg.~\ref{alg:main1} provide a solution to Prob.~\ref{prob:main}. 
\end{theorem}

The complexity of one iteration of the solution is linear with respect to the size of $\P^H$, as the applied graph search algorithms are linear (see, e.g. ~\cite{cormen}). The size of the intersection automaton $\A^h$ is in $\mathcal O (n^{|I_\ell|+1})$, where $n$ is the maximal set of states reachable in some $\B_i$, $i \in I_\ell$ within horizon $h$. The size of the product $\P^H$ is $\mathcal O(n^{|I_\ell| + 1})$, where $n$ is the maximal set of states reachable in some $\B_i$ or $\T_i$, $i\in I_\ell$ within the horizon $H$. In the worst case, when $n$ reaches the sizes of $\B_i$ or $\T_i$ respectively, and when the number of dependency classes $n_\ell = 1$, the complexity of one iteration reaches the one of the straightforward solution, i.e., the complexity of one iteration is in $\mathcal O (N\cdot \prod_{i \in N} |\T_i| \cdot |\B_i|)$, where $|\T_i|$ and $|\B_i|$ is the size of $\T_i$ and $\B_i$, respectively. However, as we demonstrate in Sec.~\ref{sec:simulations}, a dramatic improvement of computational times can be achieved in practice.

Assump.~\ref{assump:BA} may be violated for two different reasons: First, if the selected horizon $\h$ is too short, and although $F_\A = \emptyset$ in $\A^h$, there exists $\h' > h$, such that $F_\A \neq \emptyset$ in $\A^{h'}$. Second, if $F_\A = \emptyset$ even for $h \to \infty$. 
We propose to systematically extend the horizon $h$ and update the automaton $\A^h$ until a set of states $F_\A$ becomes nonempty, or until the extension does not change the automaton $\A^h$ any more. 
 In the former case, the automaton $\A^h$ with the extended horizon satisfies Assump.~\ref{assump:BA} and thus is used in constructing $\P^H$, maintaining the remainder of the solution as described in Sec.~\ref{sec:product} and \ref{sec:infinite}. In the latter case the specification has become infeasible, indicating that a wrong step has been made in past.
Therefore, we backtrack along the executed solution to a point when another service could have been executed instead of the one that has been already done. Intuitively, we ``undo'' the service and mark this service as forbidden in the specification automata. The backtracking procedure is roughly summarized in Alg.~\ref{alg:backtrack}.
In order to perform the backtracking, the system execution prefixes have to be remembered. To reduce the memory requirements, note that cycles between two exact same system execution states can be removed from the system execution prefixes without any harm.
As there are only finitely many transitions possible in each system state of each TS and each BA, the backtracking procedure will ensure that eventually, the agents' trace prefixes will be found by Alg.~\ref{alg:short} without further backtracking. 
\label{sec:relax:PA}
Once Assump.~\ref{assump:BA} holds, there is only one reason for violation of Assump.~\ref{assump:PA}, which is that the planning horizon $\H$ is not long enough. To cope with this, we systematically extend the horizon $\H$ similarly as we extended $\h$ in the BA. Eventually, a progressive state will be found.

\begin{algorithm}[!h]
\caption{Backtracking}
\label{alg:backtrack}
\begin{algorithmic}[1]
\footnotesize
\INPUT  Transition systems $\T_1,\ldots, \T_N$; BAs $\B_1,\ldots, \B_N$; \emph{System execution prefix} $(\tau_1^{\mathfrak t},\ldots,\tau_N^{\mathfrak t},\rho_1^{\mathfrak t},\ldots,\rho_N^{\mathfrak t})$ up to the current time $\mathfrak t$, where $\tau_i^{\mathfrak t} = s_{i,1}\varpi_{i,1}\ldots \varpi_{i,{\mathfrak t}-1}s_{i,\mathfrak t}$, and $\rho_i^{\mathfrak t} = \rho_{i,1} \ldots \rho_{i,\mathfrak t}$, for all $i \in \{1,\ldots N\}$.
\OUTPUT Updates to BAs $\B_1,\ldots, \B_N$
\STATE $k := \mathfrak t$
\WHILE {solution not found}
\STATE $k := k -1$
\STATE Check, if the execution of $\bigcup_{1 \in \{1,\ldots,N\}} \varpi_{i,k}$ can lead to a different set of states of BAs than to $q_{1,t},\ldots, q_{N,t}$. If so, apply the change and goto line 6.
\STATE Forbid $\bigcup_{1 \in \{1,\ldots,N\}} \varpi_{i,k}$ in the states $q_{1,k},\ldots, q_{N,k}$ of each respective automaton $\B_1,\ldots,\B_N$
\STATE Execute one iteration of  Alg.~\ref{alg:short} from $\currs_1 = s_{1,k},\ldots, \currs_N=s_{N,k}, \currq_1 = q_{1,k},\ldots, \currq_N=q_{N,k}$
\STATE If a plan was found in line 5, continue with execution of Alg.~\ref{alg:short}, otherwise goto line 2 of Backtracking.
\ENDWHILE
\end{algorithmic}
\end{algorithm}

Note, that Assump.~\ref{assump:BA} and~\ref{assump:PA} can be enforced by the selection a large enough $h$ and $H$, respectively. Particularly, $h \geq \max_{i\in N} |Q_i|$, and $H \geq \max_{i\in N}|S_i|$ ensures the completeness of our approach. However, in such a case, the complexity of the proposed approach meets the complexity of the straightforward solution in Sec~\ref{sec:pf:cs}.
A good guidance criterion for the choice of appropriate size of the receding horizon is the maximum, the average, or the mean of the shortest distance (i.e., the smallest number of transitions) between two actions labeled with non-silent service sets in the given TSs. If the selected horizon is too short, there is frequently no action labeled with a non-silent service set present in the intersection automaton, and the horizon gets frequently extended by the algorithm from Sec. 4.5.1. If the horizon is slightly longer, the resulting intersection automata contain only a few actions labeled with non-silent service sets, and the backtracking from Sec. 4.5.2 might take place quite often. On the other hand, if the selected horizon is too long, then the intersection automaton might be too large to be efficiently handled. The goal is to select a horizon to achieve a reasonable size of the intersection automaton (according to our experience, hundreds to thousands of states maximally) while containing as many actions labeled with non-silent services as possible.

\section{Example}
\label{sec:simulations}
To demonstrate our approach and its benefits, \jana{we consider the system from Example \ref{example: running}.}
We have implemented the proposed solution in MATLAB, and we illustrate snapshots of the resulting trace {under stepwise synchronization} in~Fig.~\ref{fig:example} (A)-(D). It can be seen that the agents make progress towards satisfaction of their respective formulas. 
In the computation, the default values of planning horizons were $h=3$, and $H=5$. In several cases, the latter value had to be extended as described in~Sec.\ref{sec:discussion}. The maximum value needed in order to find a solution was $H=9$.
The sizes of the product automata handled in each iteration of the algorithm have significantly reduced in comparison to the straightforward centralized solution from Sec.~\ref{sec:pf}, where all three agents belong to the dependency class yielding thus a synchronized TS with $144^3 \approx 3$ million states. With our dynamic decomposition, at most two agents belong to the same dependency class at the time, resulting into product automata sizes in order of thousands states and the computation of each iteration took seconds. When the agents are not dependent on each other within $h$ the sizes of product automata are tens to hundreds states and the computation of each iteration took seconds to minutes.
The durations of all agents' transitions were randomly generated from $\{5,...,10\}$ time units. The first 7 services in the plan of agent 2 have been completed after 54 iterations, with average duration of $\approx 477.1$ time units.
In the event-triggered solution, the individual resulting traces did not change, however, as indicated in Fig~\ref{fig:example}.(E), the randomized transition durations caused some of the agents progress more and some of them less in comparison to the stepwise solution. Average number of iterations to provide the 7  services of agent 2 was 30.3, and average time of completion was $494.2$ time units (computed from 20 simulation cases). Finally,  Fig~\ref{fig:example}.(F) shows the outcome of the event-triggered solution after $477$ time units when the duration of transitions of agent 2 was changed to a random number between in $\{1,\ldots,5\}$. The average number of iterations to provide the first 7  services of agent 2  was 26.3, and average time of completion was $355.6$ time units (from 20 simulations). The outcome of the stepwise solution for this case did not change, except for the average time of completion, which is now $\approx 457.5$. This case thus demonstrates better suitability of the event-triggered solution for heterogeneous multi-agent systems.
\begin{figure*}[!t]
\begin{center}
\includegraphics[width=0.32\linewidth]{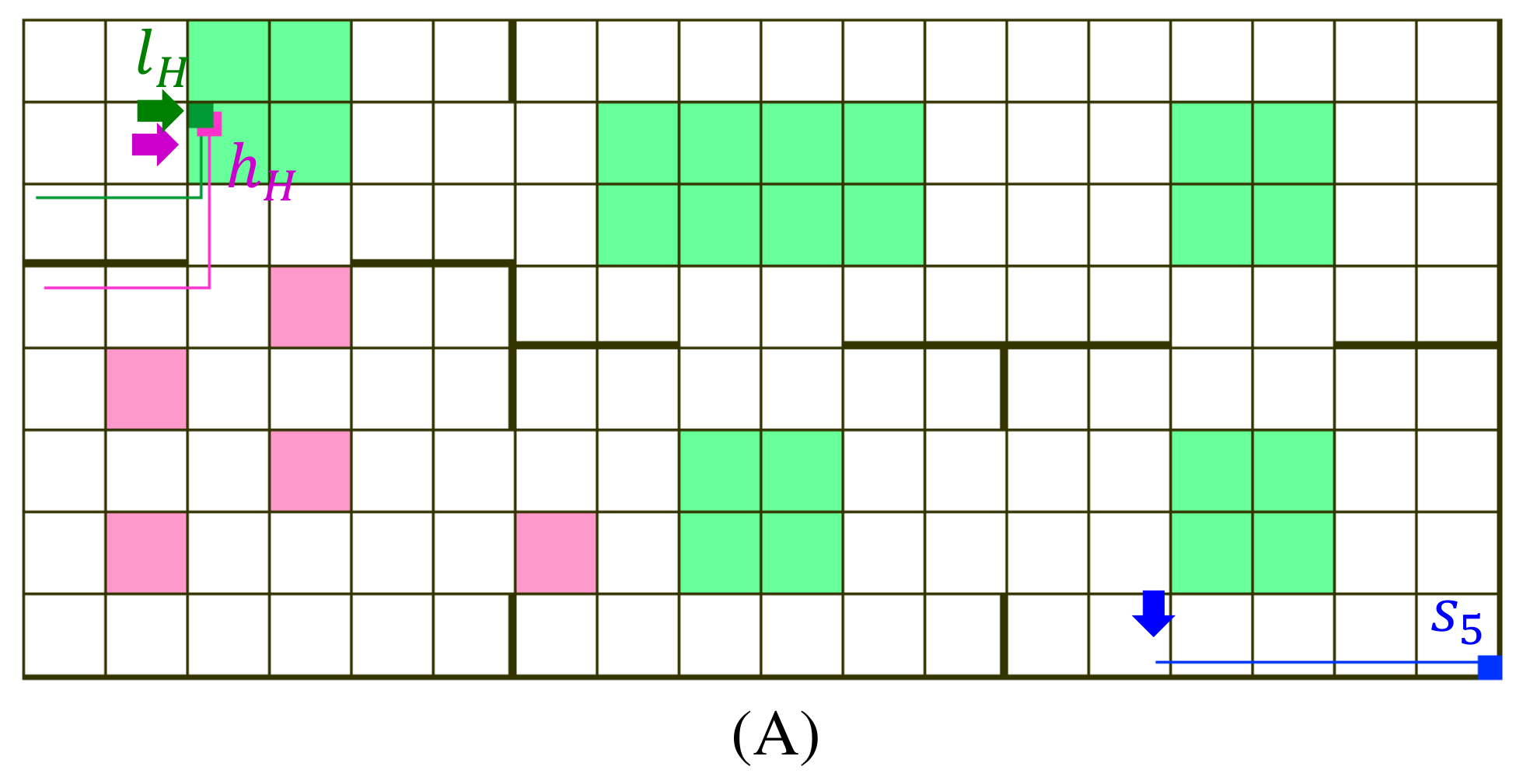} 
\includegraphics[width=0.32\linewidth]{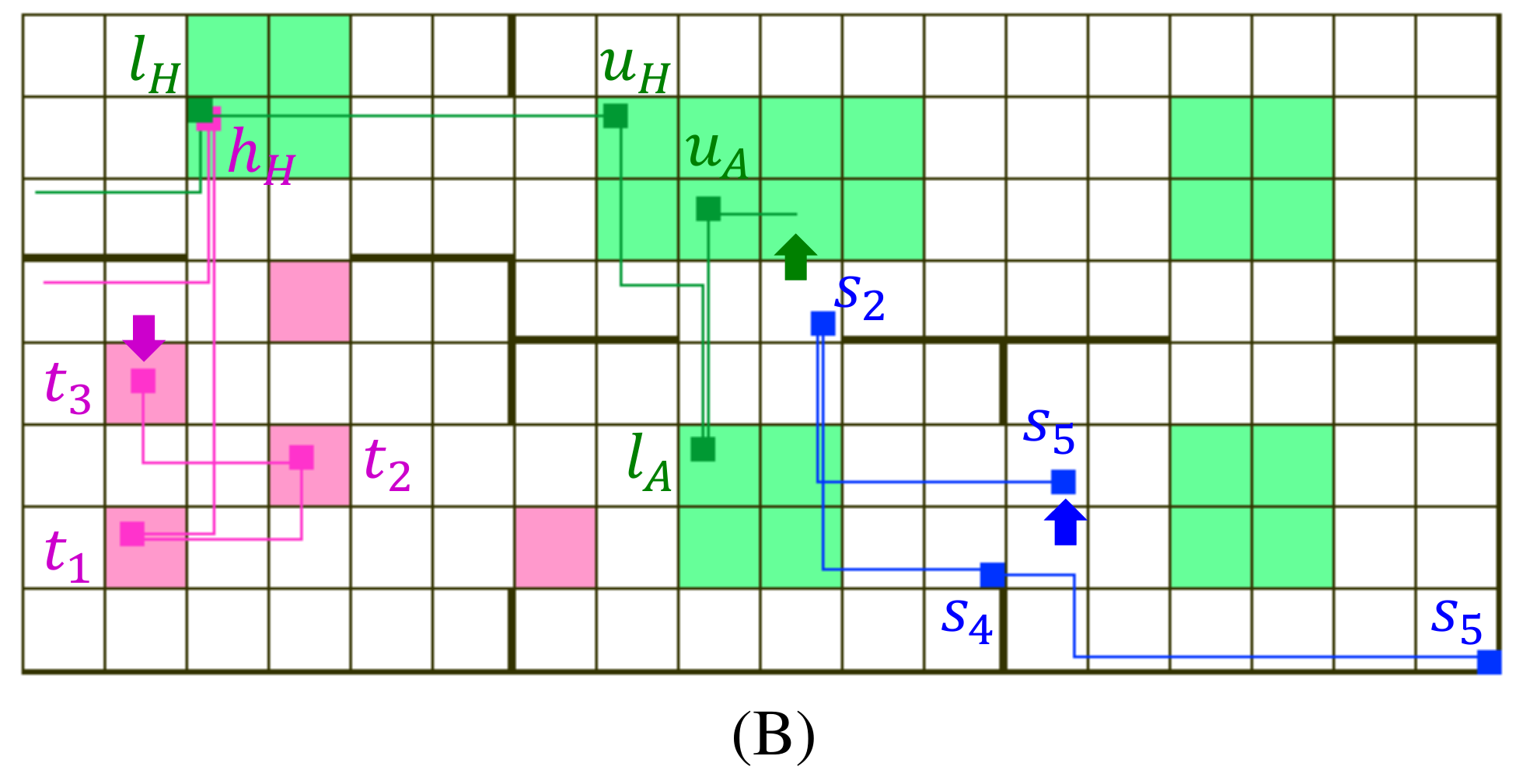} 
\includegraphics[width=0.32\linewidth]{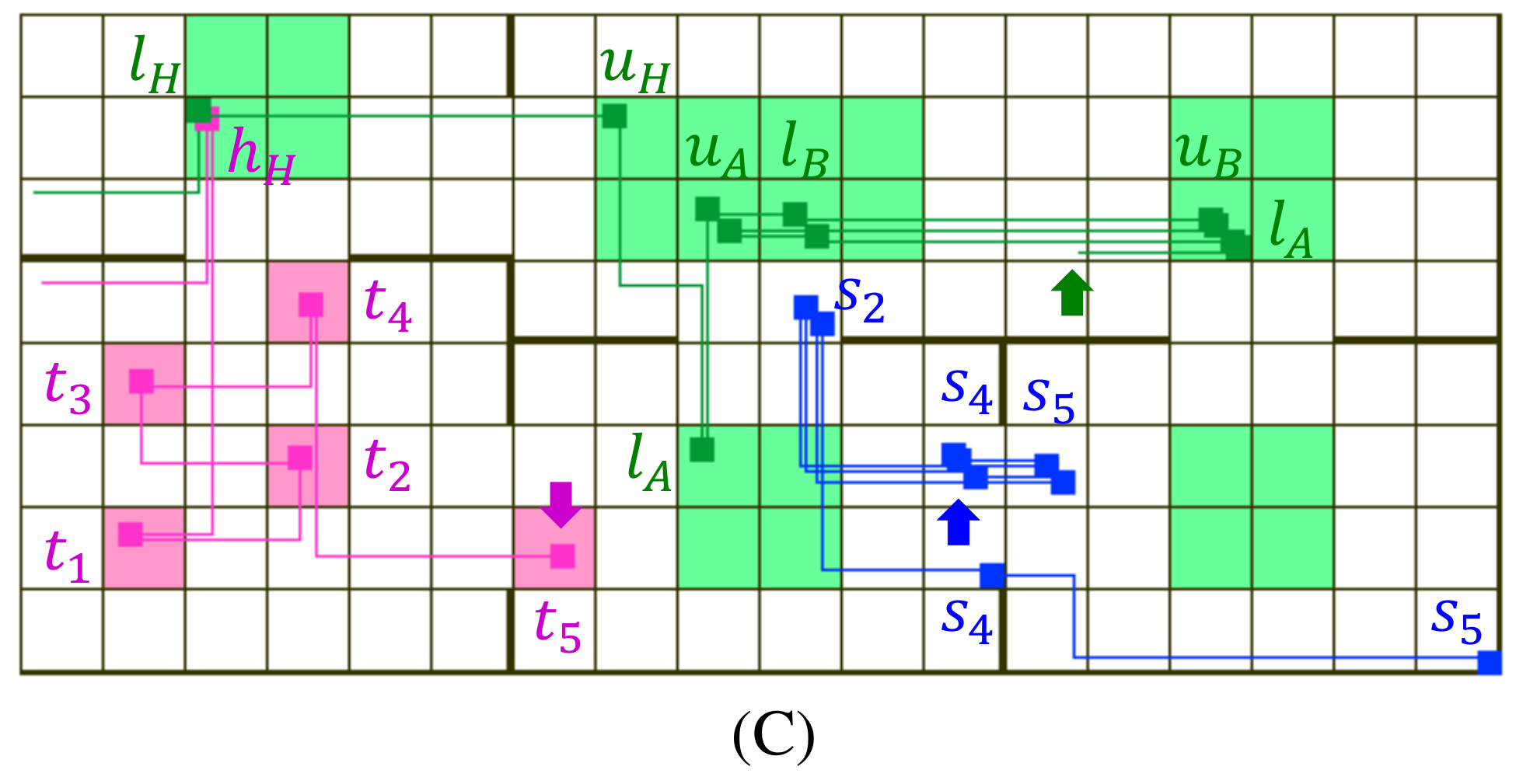}  \\ 
\includegraphics[width=0.32\linewidth]{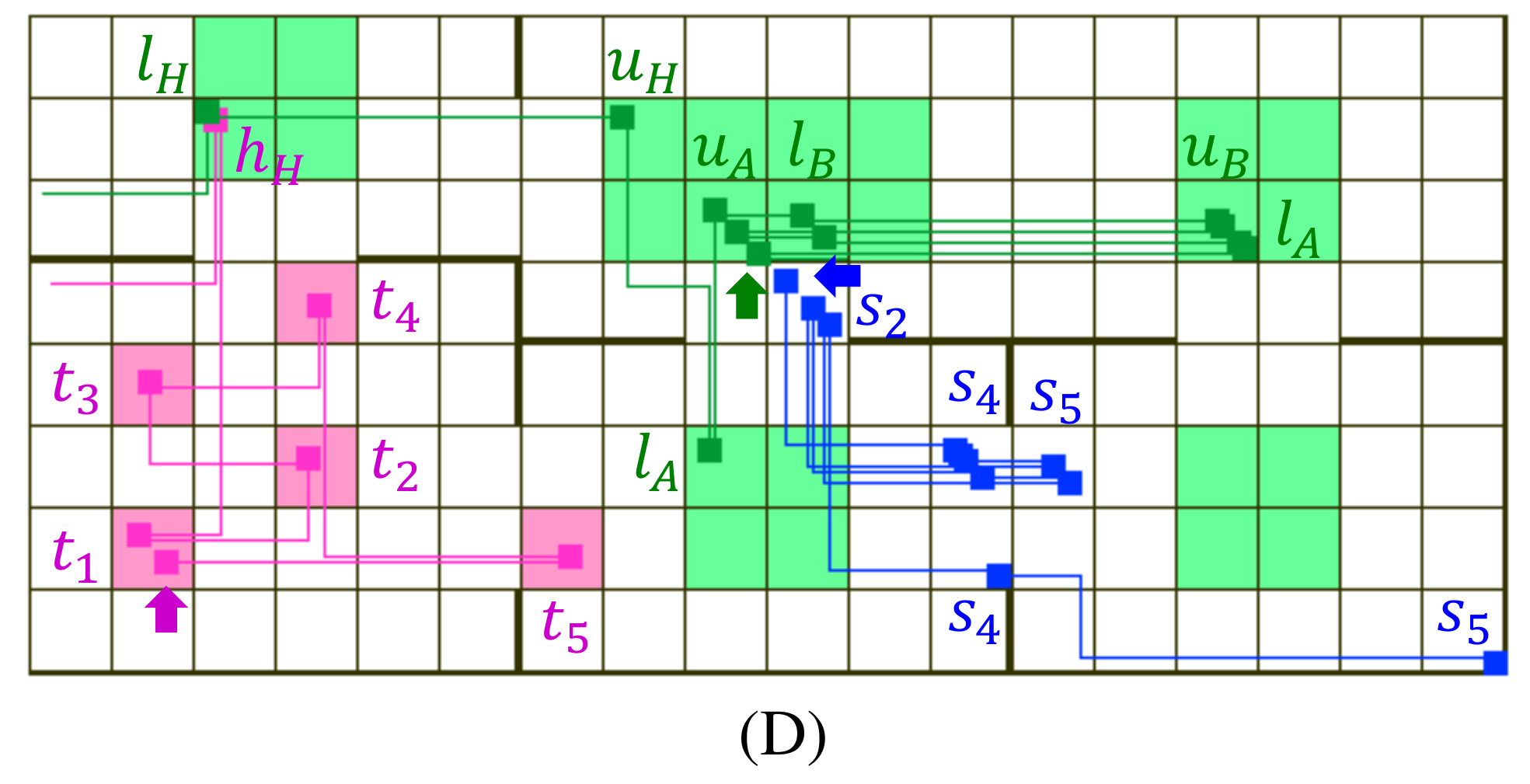} 
\includegraphics[width=0.32\linewidth]{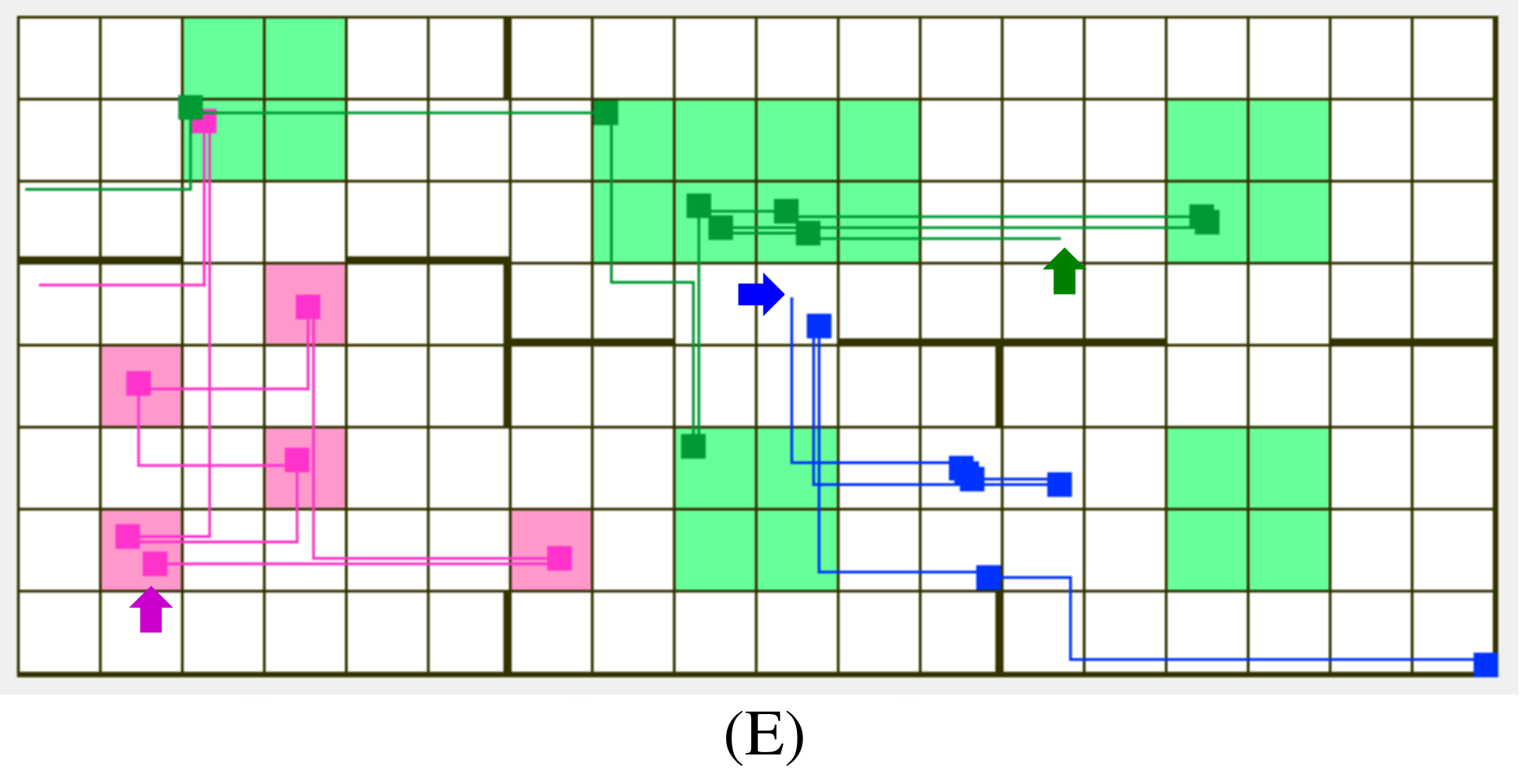}
\includegraphics[width=0.32\linewidth]{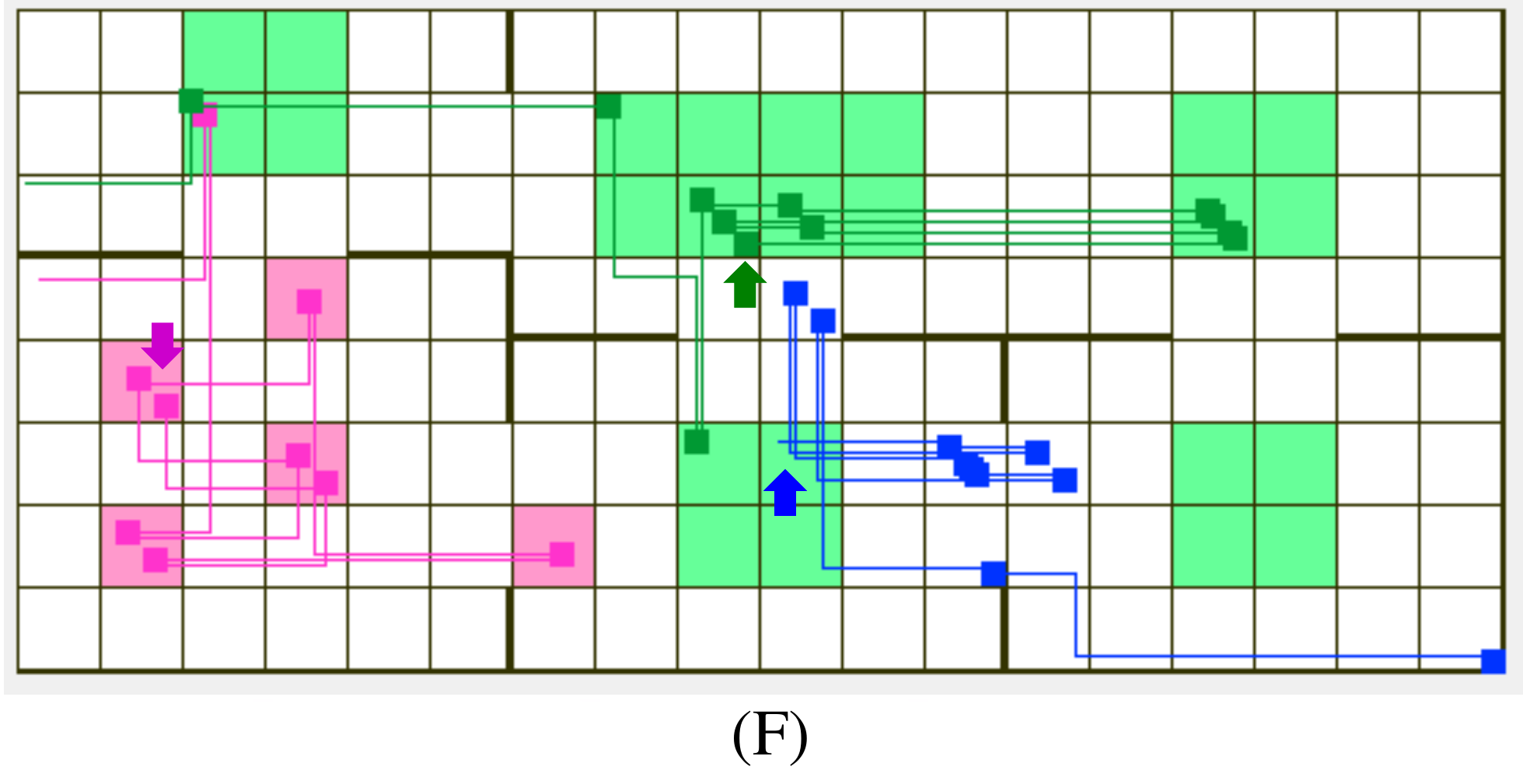}
\end{center}
\caption{\scriptsize
Trace prefixes of agent 1 (green/light), agent 2 (pink/medium), and agent 3 (blue/dark). The initial position of the agents are in the bottom left corner of $R_1$, in the top left corner of $R_3$, and in the  bottom right corner of $R_5$, respectively. The services are depicted as squares, the current position of the agents at the moment of the snapshot are indicated with arrows. In all figures, services $l_H$ and $h_H$, and $t_5$ and $s_4$ are provided at the same time. Specifically, (A) depicts the moment when $l_H$ and $h_H$ are provided and (C) depict the moment when $t_5$ and $s_4$ are provided. (A)-(D): the outcome of the stepwise solution after 1, 4, 6, and 7 services of agent 2 are provided, which is after 5, 22, 48, and 54 iterations, respectively; (E) an example outcome of the event-triggered solution after 7 services of agent 2, which is after 24 iterations; (F) an example outcome of the event-triggered solution after 477 time units in case agent 2 is faster in executing its transitions than the other two; 9 services were completed by agent 2 in 37 iterations.
}
\label{fig:example}
\end{figure*}

\section{Summary and Future Work}
\label{sec:summary}
We have proposed an automata-based receding horizon approach to solve the multi-agent planning problem from local LTL specifications. The solution decomposes the infinite horizon planning problem into finite horizon planning problems that are solved iteratively. It enables each agent to restrict its focus only on the agents that are constrained by its formula within a limited horizon, and hence to decentralize the planning procedure. Moreover, via considering the finite horizon, we reduce the size of handled state space. Stepwise synchronization can be substituted with less frequent event-based synchronization, increasing the independence of the agents during the plan execution.
Future research directions include involving various optimality requirements,
or robustness to small perturbations as well as evaluation of the approach using mobile robots.

\begin{ack}
The authors are with the ACCESS Linnaeus Center, School of Electrical
Engineering, KTH Royal Institute of Technology, SE-100 44, Stockholm,
Sweden and with the KTH Centre for
Autonomous Systems. This work was supported by the EU STREP RECONFIG, and by the H2020 ERC Starting Grant BUCOPHSYS.
\end{ack}

\bibliographystyle{plain}
\bibliography{refer}

\begin{thebibliography}{10}

\bibitem{principles}
C.~Baier and J.-P. Katoen.
\newblock {\em Principles of Model Checking}.
\newblock MIT Press, 2008.

\bibitem{Belta-TAC06}
C.~Belta and L.~C. G. J.~M. Habets.
\newblock Control of a class of nonlinear systems on rectangles.
\newblock {\em {IEEE} Transactions on Automatic Control}, 51(11):1749--1759,
  2006.

\bibitem{kavraki-ram}
A.~Bhatia, M.~R. Maly, L.~E. Kavraki, and M.~Y. Vardi.
\newblock Motion planning with complex goals.
\newblock {\em IEEE Robotics Automation Magazine}, 18(3):55 --64, 2011.

\bibitem{yushan-tr2012}
Y.~Chen, X.~C. Ding, A.~Stefanescu, and C.~Belta.
\newblock Formal approach to the deployment of distributed robotic teams.
\newblock {\em IEEE Transactions on Robotics}, 28(1):158--171, 2012.

\bibitem{cormen}
T.~H. Cormen, C.~Stein, R.~L. Rivest, and C.~E. Leiserson.
\newblock {\em Introduction to Algorithms}.
\newblock McGraw-Hill Higher Education, 2nd edition, 2001.

\bibitem{dennis-rh2}
X.~C. Ding, C.~Belta, and Cassandras~C. G.
\newblock Receding horizon surveillance with temporal logic specifications.
\newblock In {\em {IEEE} Conference on Decision and Control}, pages 256--261,
  2010.

\bibitem{dimos-cdc12}
I.~Filippidis, D.V. Dimarogonas, and K.J. Kyriakopoulos.
\newblock Decentralized multi-agent control from local {LTL} specifications.
\newblock In {\em {IEEE} Conference on Decision and Control}, pages 6235--6240,
  2012.

\bibitem{meng-ijrr2015}
M.~Guo and D.~V. Dimarogonas.
\newblock Multi-agent plan reconfiguration under local {LTL} specifications.
\newblock {\em International Journal of Robotics Research}, 34(2):218--235,
  2015.

\bibitem{hadas-icra2012}
G.~Jing, C.~Finucane, V.~Raman, and H.~Kress-Gazit.
\newblock Correct high-level robot control from structured english.
\newblock In {\em {IEEE} International Conference on Robotics and Automation},
  pages 3543--3544, 2012.

\bibitem{sertac-ijnc2010}
S.~Karaman and E.~Frazzoli.
\newblock Vehicle routing with temporal logic specifications: Applications to
  multi-{UAV} mission planning.
\newblock {\em International Journal of Robust and Nonlinear Control},
  21:1372--1395, 2011.

\bibitem{marius-tac2008}
M.~Kloetzer and C.~Belta.
\newblock {A} fully automated framework for control of linear systems from
  temporal logic specifications.
\newblock {\em {IEEE} {T}ransactions on {A}utomatic {C}ontrol}, 53(1):287--297,
  2008.

\bibitem{marius-cdc2011}
M.~Kloetzer, X.~C. Ding, and C.~Belta.
\newblock Multi-robot deployment from {LTL} specifications with reduced
  communication.
\newblock In {\em {IEEE} Conference on Decision and Control and European
  Control Conference}, pages 4867--4872, 2011.

\bibitem{hadas09TL}
H.~Kress-Gazit, G.~E. Fainekos, and G.~J. Pappas.
\newblock Temporal logic-based reactive mission and motion planning.
\newblock {\em IEEE Transactions on Robotics}, 25(6):1370--1381, 2009.

\bibitem{lavalle}
S.~M. LaValle.
\newblock {\em Planning Algorithms}.
\newblock Cambridge University Press, 2006.

\bibitem{loizou-cdc2005}
S.~G. Loizou and K.~J. Kyriakopoulos.
\newblock Automated planning of motion tasks for multi-robot systems.
\newblock In {\em {IEEE} Conference on Decision and Control}, pages 78 -- 83,
  2005.

\bibitem{quo-icra2004}
M.M. Quottrup, T.~Bak, and R.I. Zamanabadi.
\newblock Multi-robot planning : a timed automata approach.
\newblock In {\em {IEEE} International Conference on Robotics and Automation},
  volume~5, pages 4417--4422, 2004.

\bibitem{maja}
M.~Svorenova, J.~Tumova, J.~Barnat, and I.~Cerna.
\newblock Attraction-based receding horizon path planning with temporal logic
  constraints.
\newblock In {\em {IEEE} Conference on Decision and Control}, pages 6749--6754,
  2012.

\bibitem{acc14}
J.~Tumova and D.~Dimarogonas.
\newblock A receding horizon approach to multi-agent planning from local {LTL}
  specifications.
\newblock In {\em American Control Conference}, pages 1775--1780, 2014.

\bibitem{alphan-ijrr2013}
A.~Ulusoy, S.~L. Smith, X.~C. Ding, C.~Belta, and D.~Rus.
\newblock Optimality and robustness in multi-robot path planning with temporal
  logic constraints.
\newblock {\em International Journal of Robotics Research}, 32(8):889--911,
  2013.

\bibitem{lygeros-ecc2013}
C.~Wiltsche, F.~A. Ramponi, and J.~Lygeros.
\newblock Synthesis of an asynchronous communication protocol for search and
  rescue robots.
\newblock In {\em European Control Conference}, pages 1256--1261, 2013.

\bibitem{nok-hscc2010}
T.~Wongpiromsarn, U.~Topcu, and R.~M. Murray.
\newblock {R}eceding horizon control for temporal logic specifications.
\newblock In {\em Hybrid systems: Computation and Control}, pages 101--110,
  2010.

\end{thebibliography}

\end{document}